\documentclass[traditabstract,longauth]{aa}
\usepackage{amssymb,hyperref,units,natbib,graphicx,txfonts}
\usepackage[switch]{lineno}
\usepackage{breakurl} 
\newcommand{\un}[1]{\,\mathrm{#1}}
\newcommand{\nue}{\ensuremath{\nu_{\rm e}}}

\newcommand{\degC}[1]{\mbox{$#1\,^\circ{\mathrm C}$}}
\newcommand{\inch}{\texttt{"}} 
\begin{document}
\title{IceCube Sensitivity for Low-Energy Neutrinos \\from Nearby Supernovae}
\author{\normalsize
IceCube Collaboration:
R.~Abbasi\inst{\ref{r1}}\and
Y.~Abdou\inst{\ref{r2}}\and
T.~Abu-Zayyad\inst{\ref{r3}}\and
M.~Ackermann\inst{\ref{r10}}\and
J.~Adams\inst{\ref{r4}}\and
J.~A.~Aguilar\inst{\ref{r1}}\and
M.~Ahlers\inst{\ref{r5}}\and
M.~M.~Allen\inst{\ref{r26}}\and
D.~Altmann\inst{\ref{r18}}\and
K.~Andeen\inst{\ref{r1}}\inst{,}\inst{\ref{r44}}\and
J.~Auffenberg\inst{\ref{r6}}\and
X.~Bai\inst{\ref{r7}}\inst{,}\inst{\ref{r45}}\and
M.~Baker\inst{\ref{r1}}\and
S.~W.~Barwick\inst{\ref{r8}}\and
V.~Baum\inst{\ref{r33}}\and
R.~Bay\inst{\ref{r9}}\and
J.~L.~Bazo~Alba\inst{\ref{r10}}\and
K.~Beattie\inst{\ref{r11}}\and
J.~J.~Beatty\inst{\ref{r12}}\inst{,}\inst{\ref{r13}}\and
S.~Bechet\inst{\ref{r14}}\and
J.~K.~Becker\inst{\ref{r15}}\and
K.-H.~Becker\inst{\ref{r6}}\and
M.~L.~Benabderrahmane\inst{\ref{r10}}\and
S.~BenZvi\inst{\ref{r1}}\and
J.~Berdermann\inst{\ref{r10}}\and
P.~Berghaus\inst{\ref{r1}}\and
D.~Berley\inst{\ref{r16}}\and
E.~Bernardini\inst{\ref{r10}}\and
D.~Bertrand\inst{\ref{r14}}\and
D.~Z.~Besson\inst{\ref{r17}}\and
D.~Bindig\inst{\ref{r6}}\and
M.~Bissok\inst{\ref{r18}}\and
E.~Blaufuss\inst{\ref{r16}}\and
J.~Blumenthal\inst{\ref{r18}}\and
D.~J.~Boersma\inst{\ref{r18}}\and
C.~Bohm\inst{\ref{r19}}\and
D.~Bose\inst{\ref{r20}}\and
S.~B\"oser\inst{\ref{r21}}\and
O.~Botner\inst{\ref{r22}}\and
A.~M.~Brown\inst{\ref{r4}}\and
S.~Buitink\inst{\ref{r11}}\and
K.~S.~Caballero-Mora\inst{\ref{r26}}\and
M.~Carson\inst{\ref{r2}}\and
D.~Chirkin\inst{\ref{r1}}\and
B.~Christy\inst{\ref{r16}}\and
F.~Clevermann\inst{\ref{r23}}\and
S.~Cohen\inst{\ref{r24}}\and
C.~Colnard\inst{\ref{r25}}\and
D.~F.~Cowen\inst{\ref{r26}}\inst{,}\inst{\ref{r27}}\and
A.~H.~Cruz Silva\inst{\ref{r10}}\and
M.~V.~D'Agostino\inst{\ref{r9}}\and
M.~Danninger\inst{\ref{r19}}\and
J.~Daughhetee\inst{\ref{r28}}\and
J.~C.~Davis\inst{\ref{r12}}\and
C.~De~Clercq\inst{\ref{r20}}\and
T.~Degner\inst{\ref{r21}}\and
L.~Demir\"ors\inst{\ref{r24}}\and
F.~Descamps\inst{\ref{r2}}\and
P.~Desiati\inst{\ref{r1}}\and
G.~de~Vries-Uiterweerd\inst{\ref{r2}}\and
T.~DeYoung\inst{\ref{r26}}\and
J.~C.~D{\'\i}az-V\'elez\inst{\ref{r1}}\and
M.~Dierckxsens\inst{\ref{r14}}\and
J.~Dreyer\inst{\ref{r15}}\and
J.~P.~Dumm\inst{\ref{r1}}\and
M.~Dunkman\inst{\ref{r26}}\and
J.~Eisch\inst{\ref{r1}}\and
R.~W.~Ellsworth\inst{\ref{r16}}\and
O.~Engdeg{\aa}rd\inst{\ref{r22}}\and
S.~Euler\inst{\ref{r18}}\and
P.~A.~Evenson\inst{\ref{r7}}\and
O.~Fadiran\inst{\ref{r29}}\and
A.~R.~Fazely\inst{\ref{r30}}\and
A.~Fedynitch\inst{\ref{r15}}\and
J.~Feintzeig\inst{\ref{r1}}\and
T.~Feusels\inst{\ref{r2}}\and
K.~Filimonov\inst{\ref{r9}}\and
C.~Finley\inst{\ref{r19}}\and
T.~Fischer-Wasels\inst{\ref{r6}}\and
B.~D.~Fox\inst{\ref{r26}}\and
A.~Franckowiak\inst{\ref{r21}}\and
R.~Franke\inst{\ref{r10}}\and
T.~K.~Gaisser\inst{\ref{r7}}\and
J.~Gallagher\inst{\ref{r31}}\and
L.~Gerhardt\inst{\ref{r11}}\inst{,}\inst{\ref{r9}}\and
L.~Gladstone\inst{\ref{r1}}\and
T.~Gl\"usenkamp\inst{\ref{r18}}\and
A.~Goldschmidt\inst{\ref{r11}}\and
J.~A.~Goodman\inst{\ref{r16}}\and
D.~G\'ora\inst{\ref{r10}}\and
D.~Grant\inst{\ref{r32}}\and
T.~Griesel\inst{\ref{r33}}\and
A.~Gro{\ss}\inst{\ref{r4}}\inst{,}\inst{\ref{r25}}\and
S.~Grullon\inst{\ref{r1}}\and
M.~Gurtner\inst{\ref{r6}}\and
C.~Ha\inst{\ref{r26}}\and
A.~Haj Ismail\inst{\ref{r2}}\and
A.~Hallgren\inst{\ref{r22}}\and
F.~Halzen\inst{\ref{r1}}\and
K.~Han\inst{\ref{r4}}\and
K.~Hanson\inst{\ref{r14}}\inst{,}\inst{\ref{r1}}\and
D.~Heinen\inst{\ref{r18}}\and
K.~Helbing\inst{\ref{r6}}\and
R.~Hellauer\inst{\ref{r16}}\and
S.~Hickford\inst{\ref{r4}}\and
G.~C.~Hill\inst{\ref{r1}}\and
K.~D.~Hoffman\inst{\ref{r16}}\and 
B.~Hoffmann\inst{\ref{r18}}\and 
A.~Homeier\inst{\ref{r21}}\and
K.~Hoshina\inst{\ref{r1}}\and
W.~Huelsnitz\inst{\ref{r16}}\inst{,}\inst{\ref{r46}}\and
J.-P.~H\"ul{\ss}\inst{\ref{r18}}\and
P.~O.~Hulth\inst{\ref{r19}}\and
K.~Hultqvist\inst{\ref{r19}}\and
S.~Hussain\inst{\ref{r7}}\and
A.~Ishihara\inst{\ref{r35}}\and
E.~Jakobi\inst{\ref{r10}}\and 
J.~Jacobsen\inst{\ref{r1}}\and
G.~S.~Japaridze\inst{\ref{r29}}\and
H.~Johansson\inst{\ref{r19}}\and
K.-H.~Kampert\inst{\ref{r6}}\and
A.~Kappes\inst{\ref{r36}}\and
T.~Karg\inst{\ref{r6}}\and
A.~Karle\inst{\ref{r1}}\and
P.~Kenny\inst{\ref{r17}}\and
J.~Kiryluk\inst{\ref{r11}}\inst{,}\inst{\ref{r9}}\and
F.~Kislat\inst{\ref{r10}}\and
S.~R.~Klein\inst{\ref{r11}}\inst{,}\inst{\ref{r9}}\and
H.~K\"ohne\inst{\ref{r23}}\and
G.~Kohnen\inst{\ref{r34}}\and
H.~Kolanoski\inst{\ref{r36}}\and
L.~K\"opke\inst{\ref{r33}}\and
S.~Kopper\inst{\ref{r6}}\and
D.~J.~Koskinen\inst{\ref{r26}}\and
M.~Kowalski\inst{\ref{r21}}\and
T.~Kowarik\inst{\ref{r33}}\and
M.~Krasberg\inst{\ref{r1}}\and
G.~Kroll\inst{\ref{r33}}\and
N.~Kurahashi\inst{\ref{r1}}\and 
T.~Kuwabara\inst{\ref{r7}}\and
M.~Labare\inst{\ref{r20}}\and
K.~Laihem\inst{\ref{r18}}\and
H.~Landsman\inst{\ref{r1}}\and
M.~J.~Larson\inst{\ref{r26}}\and
R.~Lauer\inst{\ref{r10}}\and
J.~L\"unemann\inst{\ref{r33}}\and
J.~Madsen\inst{\ref{r3}}\and
A.~Marotta\inst{\ref{r14}}\and
R.~Maruyama\inst{\ref{r1}}\and
K.~Mase\inst{\ref{r35}}\and
H.~S.~Matis\inst{\ref{r11}}\and
K.~Meagher\inst{\ref{r16}}\and
M.~Merck\inst{\ref{r1}}\and
P.~M\'esz\'aros\inst{\ref{r27}}\inst{,}\inst{\ref{r26}}\and
T.~Meures\inst{\ref{r18}}\and
S.~Miarecki\inst{\ref{r11}}\inst{,}\inst{\ref{r9}}\and 
E.~Middell\inst{\ref{r10}}\and
N.~Milke\inst{\ref{r23}}\and
J.~Miller\inst{\ref{r22}}\and
T.~Montaruli\inst{\ref{r1}}\inst{,}\inst{\ref{r37}}\and
R.~Morse\inst{\ref{r1}}\and
S.~M.~Movit\inst{\ref{r27}}\and
R.~Nahnhauer\inst{\ref{r10}}\and
J.~W.~Nam\inst{\ref{r8}}\and
U.~Naumann\inst{\ref{r6}}\and
D.~R.~Nygren\inst{\ref{r11}}\and
S.~Odrowski\inst{\ref{r25}}\and
A.~Olivas\inst{\ref{r16}}\and
M.~Olivo\inst{\ref{r22}}\inst{,}\inst{\ref{r15}}\and
A.~O'Murchadha\inst{\ref{r1}}\and
S.~Panknin\inst{\ref{r21}}\and
L.~Paul\inst{\ref{r18}}\and
C.~P\'erez~de~los~Heros\inst{\ref{r22}}\and
J.~Petrovic\inst{\ref{r14}}\and
A.~Piegsa\inst{\ref{r33}}\and
D.~Pieloth\inst{\ref{r23}}\and
R.~Porrata\inst{\ref{r9}}\and
J.~Posselt\inst{\ref{r6}}\and
P.~B.~Price\inst{\ref{r9}}\and
G.~T.~Przybylski\inst{\ref{r11}}\and
K.~Rawlins\inst{\ref{r38}}\and
P.~Redl\inst{\ref{r16}}\and
E.~Resconi\inst{\ref{r25}}\inst{,}\inst{\ref{r42}}\and
W.~Rhode\inst{\ref{r23}}\and
M.~Ribordy\inst{\ref{r24}}\and
A.~S.~Richard\inst{\ref{r30}}\and
M.~Richman\inst{\ref{r16}}\and
J.~P.~Rodrigues\inst{\ref{r1}}\and
F.~Rothmaier\inst{\ref{r33}}\and
C.~Rott\inst{\ref{r12}}\and
T.~Ruhe\inst{\ref{r23}}\and
D.~Rutledge\inst{\ref{r26}}\and
B.~Ruzybayev\inst{\ref{r7}}\and
D.~Ryckbosch\inst{\ref{r2}}\and
H.-G.~Sander\inst{\ref{r33}}\and
M.~Santander\inst{\ref{r1}}\and
S.~Sarkar\inst{\ref{r5}}\and
K.~Schatto\inst{\ref{r33}}\and
T.~Schmidt\inst{\ref{r16}}\and
A.~Sch\"onwald\inst{\ref{r10}}\and
A.~Schukraft\inst{\ref{r18}}\and
L.~Schulte\inst{\ref{r33}}\and
A.~Schultes\inst{\ref{r6}}\and
O.~Schulz\inst{\ref{r25}}\inst{,}\inst{\ref{r43}}\and
M.~Schunck\inst{\ref{r18}}\and
D.~Seckel\inst{\ref{r7}}\and
B.~Semburg\inst{\ref{r6}}\and
S.~H.~Seo\inst{\ref{r19}}\and
Y.~Sestayo\inst{\ref{r25}}\and
S.~Seunarine\inst{\ref{r39}}\and
A.~Silvestri\inst{\ref{r8}}\and
K.~Singh\inst{\ref{r20}}\and
A.~Slipak\inst{\ref{r26}}\and
G.~M.~Spiczak\inst{\ref{r3}}\and
C.~Spiering\inst{\ref{r10}}\and
M.~Stamatikos\inst{\ref{r12}}\inst{,}\inst{\ref{r40}}\and
T.~Stanev\inst{\ref{r7}}\and
T.~Stezelberger\inst{\ref{r11}}\and
R.~G.~Stokstad\inst{\ref{r11}}\and
A.~St\"o{\ss}l\inst{\ref{r10}}\and
E.~A.~Strahler\inst{\ref{r20}}\and
R.~Str\"om\inst{\ref{r22}}\and
M.~St\"uer\inst{\ref{r21}}\and
G.~W.~Sullivan\inst{\ref{r16}}\and
Q.~Swillens\inst{\ref{r14}}\and
H.~Taavola\inst{\ref{r22}}\and
I.~Taboada\inst{\ref{r28}}\and
A.~Tamburro\inst{\ref{r3}}\and
A.~Tepe\inst{\ref{r28}}\and
S.~Ter-Antonyan\inst{\ref{r30}}\and
S.~Tilav\inst{\ref{r7}}\and
P.~A.~Toale\inst{\ref{r26}}\and
S.~Toscano\inst{\ref{r1}}\and
D.~Tosi\inst{\ref{r10}}\and
N.~van~Eijndhoven\inst{\ref{r20}}\and
J.~Vandenbroucke\inst{\ref{r9}}\and
A.~Van~Overloop\inst{\ref{r2}}\and
J.~van~Santen\inst{\ref{r1}}\and
M.~Vehring\inst{\ref{r18}}\and
M.~Voge\inst{\ref{r25}}\and
C.~Walck\inst{\ref{r19}}\and
T.~Waldenmaier\inst{\ref{r36}}\and
M.~Wallraff\inst{\ref{r18}}\and
M.~Walter\inst{\ref{r10}}\and
Ch.~Weaver\inst{\ref{r1}}\and
C.~Wendt\inst{\ref{r1}}\and
S.~Westerhoff\inst{\ref{r1}}\and
N.~Whitehorn\inst{\ref{r1}}\and
K.~Wiebe\inst{\ref{r33}}\and
C.~H.~Wiebusch\inst{\ref{r18}}\and
D.~R.~Williams\inst{\ref{r41}}\and
R.~Wischnewski\inst{\ref{r10}}\and
H.~Wissing\inst{\ref{r16}}\and
M.~Wolf\inst{\ref{r25}}\and
T.~R.~Wood\inst{\ref{r32}}\and
K.~Woschnagg\inst{\ref{r9}}\and
C.~Xu\inst{\ref{r7}}\and
D.~L.~Xu\inst{\ref{r41}}\and
X.~W.~Xu\inst{\ref{r30}}\and
J.~P.~Yanez\inst{\ref{r10}}\and
G.~Yodh\inst{\ref{r8}}\and
S.~Yoshida\inst{\ref{r35}}\and
P.~Zarzhitsky\inst{\ref{r41}}\and
M.~Zoll\inst{\ref{r19}}}
\institute{\tiny Dept.~of Physics, University of Wisconsin, Madison, WI 53706, USA\inst{\label{r1}}\and
Dept.~of Subatomic and Radiation Physics, University of Gent, B-9000 Gent, Belgium\inst{\label{r2}}\and
Dept.~of Physics, University of Wisconsin, River Falls, WI 54022, USA\inst{\label{r3}}\and
Dept.~of Physics and Astronomy, University of Canterbury, Private Bag 4800, Christchurch, New Zealand\inst{\label{r4}}\and
Dept.~of Physics, University of Oxford, 1 Keble Road, Oxford OX1 3NP, UK\inst{\label{r5}}\and
Dept.~of Physics, University of Wuppertal, D-42119 Wuppertal, Germany\inst{\label{r6}}\and
Bartol Research Institute and Department of Physics and Astronomy, University of Delaware, Newark, DE 19716, USA\inst{\label{r7}}\and
Dept.~of Physics and Astronomy, University of California, Irvine, CA 92697, USA\inst{\label{r8}}\and
Dept.~of Physics, University of California, Berkeley, CA 94720, USA\inst{\label{r9}}\and
DESY, D-15735 Zeuthen, Germany\inst{\label{r10}}\and
Lawrence Berkeley National Laboratory, Berkeley, CA 94720, USA\inst{\label{r11}}\and
Dept.~of Physics and Center for Cosmology and Astro-Particle Physics, Ohio State University, Columbus, OH 43210, USA\inst{\label{r12}}\and
Dept.~of Astronomy, Ohio State University, Columbus, OH 43210, USA\inst{\label{r13}}\and
Universit\'e Libre de Bruxelles, Science Faculty CP230, B-1050 Brussels, Belgium\inst{\label{r14}}\and
Fakult\"at f\"ur Physik \& Astronomie, Ruhr-Universit\"at Bochum, D-44780 Bochum, Germany\inst{\label{r15}}\and
Dept.~of Physics, University of Maryland, College Park, MD 20742, USA\inst{\label{r16}}\and
Dept.~of Physics and Astronomy, University of Kansas, Lawrence, KS 66045, USA\inst{\label{r17}}\and
III. Physikalisches Institut, RWTH Aachen University, D-52056 Aachen, Germany\inst{\label{r18}}\and
Oskar Klein Centre and Dept.~of Physics, Stockholm University, SE-10691 Stockholm, Sweden\inst{\label{r19}}\and
Vrije Universiteit Brussel, Dienst ELEM, B-1050 Brussels, Belgium\inst{\label{r20}}\and
Physikalisches Institut, Universit\"at Bonn, Nussallee 12, D-53115 Bonn, Germany\inst{\label{r21}}\and
Dept.~of Physics and Astronomy, Uppsala University, Box 516, S-75120 Uppsala, Sweden\inst{\label{r22}}\and
Dept.~of Physics, TU Dortmund University, D-44221 Dortmund, Germany\inst{\label{r23}}\and
Laboratory for High Energy Physics, \'Ecole Polytechnique F\'ed\'erale, CH-1015 Lausanne, Switzerland\inst{\label{r24}}\and
Max-Planck-Institut f\"ur Kernphysik, D-69177 Heidelberg, Germany\inst{\label{r25}}\and
Dept.~of Physics, Pennsylvania State University, University Park, PA 16802, USA\inst{\label{r26}}\and
Dept.~of Astronomy and Astrophysics, Pennsylvania State University, University Park, PA 16802, USA\inst{\label{r27}}\and
School of Physics and Center for Relativistic Astrophysics, Georgia Institute of Technology, Atlanta, GA 30332, USA\inst{\label{r28}}\and
CTSPS, Clark-Atlanta University, Atlanta, GA 30314, USA\inst{\label{r29}}\and
Dept.~of Physics, Southern University, Baton Rouge, LA 70813, USA\inst{\label{r30}}\and
Dept.~of Astronomy, University of Wisconsin, Madison, WI 53706, USA\inst{\label{r31}}\and
Dept.~of Physics, University of Alberta, Edmonton, Alberta, Canada T6G 2G7\inst{\label{r32}}\and 
Institute of Physics, University of Mainz, Staudinger Weg 7, D-55099 Mainz, Germany\inst{\label{r33}}\and
Universit\'e de Mons, 7000 Mons, Belgium\inst{\label{r34}}\and
Dept.~of Physics, Chiba University, Chiba 263-8522, Japan\inst{\label{r35}}\and
Institut f\"ur Physik, Humboldt-Universit\"at zu Berlin, D-12489 Berlin, Germany\inst{\label{r36}}\and
also Sezione INFN, Dipartimento di Fisica, I-70126, Bari, Italy\inst{\label{r37}}\and
Dept.~of Physics and Astronomy, University of Alaska Anchorage, 3211 Providence Dr., Anchorage, AK 99508, USA\inst{\label{r38}}\and
Dept.~of Physics, University of the West Indies, Cave Hill Campus, Bridgetown BB11000, Barbados\inst{\label{r39}}\and
NASA Goddard Space Flight Center, Greenbelt, MD 20771, USA\inst{\label{r40}}\and
Dept.~of Physics and Astronomy, University of Alabama, Tuscaloosa, AL 35487, USA\inst{\label{r41}}\and
now at T.U. Munich, 85748 Garching \& Friedrich-Alexander Universit\"at Erlangen-N\"urnberg, 91058 Erlangen, Germany\inst{\label{r42}}\and
now at T.U. Munich, 85748 Garching \inst{\label{r43}}\and
now at Dept. of Physics and Astronomy, Rutgers University, Piscataway, NJ 08854, USA\inst{\label{r44}}\and
now at Physics Department, South Dakota School of Mines and Technology, Rapid City, SD 57701, USA\inst{\label{r45}}\and
Los Alamos National Laboratory, Los Alamos, NM 87545, USA\inst{\label{r46}}
}
\date{Received  / Accepted \vspace{-10pt}}
\abstract{\vspace{-5pt}
This paper describes the response of the IceCube neutrino telescope located at the geographic South Pole to outbursts of MeV neutrinos from the core collapse of nearby massive stars. 
IceCube was completed in December 2010
forming a lattice of 5160 photomultiplier tubes that monitor a volume of $\sim$\,\unit[1]{km$^3$} in the deep Antarctic ice for particle induced photons. The telescope was designed to detect neutrinos with energies greater than 100 GeV. Owing to subfreezing ice temperatures, 
the photomultiplier dark noise rates are particularly low. Hence IceCube can also detect large numbers of MeV neutrinos by observing a collective rise in all photomultiplier rates on top of the dark noise. %
With 2 ms timing resolution, IceCube can detect subtle features in the 
temporal development of the supernova neutrino burst. For a supernova at the galactic center, its sensitivity matches that of a background-free megaton-scale supernova search experiment. The sensitivity decreases to 20 standard deviations at the galactic edge (30 kpc) and 6 standard deviations at the Large Magellanic Cloud (50 kpc). IceCube is sending triggers from potential supernovae to the Supernova Early Warning System. 
The sensitivity to neutrino properties such as the neutrino hierarchy is discussed, as well as the possibility to detect the neutronization burst, a short outbreak of 
\nue's released by electron capture on protons soon after collapse.
Tantalizing signatures, such as the formation of a quark star or a black hole as well as the characteristics of shock waves, are investigated to illustrate IceCube's capability for supernova detection. 
}
\maketitle
\section{Introduction}\label{sec:intro}
On February 23, 1987, a burst of mainly electron anti-neutrinos with energies of a few tens of MeV emitted by the Supernova SN1987A was recorded simultaneously by the Baksan~\citep{JExpTheorPhys.45.589}, IMB \citep{PhysRevLett.58.1494}, and Kamiokande-II \citep{PhysRevLett.58.1490, PhysRevD.38.448} detectors, a few hours before its optical counterpart was discovered. With just 
24 neutrinos collected, stringent limits on the mass of the $\bar\nu_{e}$, its lifetime, its magnetic moment and the number of leptonic flavors could be derived \citep{kotake}. As of now, SN1987A remains the only source of neutrinos that has been detected outside of our solar system. 
Although the optical detection of supernova explosions has a long history, detailed features of the gravitational collapse can only be studied with neutrinos, which carry away nearly 99\,\% of the gravitational binding energy soon after the collapse. The current generation of detectors is capable of detecting many orders of magnitude more neutrinos and thus it can study details of the gravitational collapse and neutrino properties.

The rate of galactic stellar collapses, including those obscured in the optical, is estimated by various methods to be $\approx (1 - 7)/ 100$ years~\citep{Diehl, Strom}. A compilation in \citet{snrate} narrows the expected range to $(1.7 - 2.5)/ 100$ years by taking into account experimental and theoretical limits. The best experimental upper limit is $<\,$\unit[9.3]/100 years~\citep{Novoseltseva}. 

While differences in the onset of the neutrino emission between various models are small \citep{PhysRevD.71.063003}, the models have yet to overcome problems with the supernova explosion mechanisms.The theoretical knowledge about the neutrino emission at times longer than several \unit[100]{$\mu$s} after the deleptonization~\citep{PhysRevLett.90.241101, AstrAstrop.450.1} is limited. However, three characteristic phases are expected: a rapid luminosity increase during collapse with the appearance of a shock breakout burst, an accretion phase ending after $\mathcal{O}(0.5)\un{s}$ during which the neutrino flux of all flavors is maximal, and a cooling phase.  The $\mathcal{O}(20)\un{s}$ duration of supernova neutrino emission is determined by the neutrino diffusion time scale in the dense matter inside the proto-neutron star. The exact features will depend on the progenitor mass with modulations introduced by the dynamics of the collapse. 

IceCube
is primarily designed to observe TeV neutrino sources with a 
wide lattice of light sensors embedded in highly transparent glacier ice used as Cherenkov medium. However, it was recognized early by \citet{Pryor} and \citet{PhysRevD.53.7359} that neutrino telescopes offer the possibility to monitor our Galaxy for supernovae.
In spite of the much lower neutrino energies of $\mathcal{O}(10\un{MeV})$ involved in a supernova burst, Cherenkov light induced by neutrino interactions will increase the count rate of all light sensors above their average value. 
Although the increase in the noise rate in each light sensor is not statistically significant, the effect will be clearly seen once the rise is considered collectively over many sensors.
Low photomultiplier noise rates, low photon absorption in the Cherenkov medium and a large number of sensors are essential. 

IceCube is uniquely suited for this measurement due to its location and 1 km$^3$ size. The noise rates in IceCube's photomultiplier tubes average around 540 Hz since they are surrounded by inert and cold ice with depth dependent temperatures ranging from \degC{-43} to \degC{-20}. At depths between \unit[(1450 -- 2450)]{m} they are also largely shielded from cosmic rays. The noise rate is further reduced by the use of detector components with reduced radioactivity. 
The detected signal rate is essentially independent of the photon scattering length and depends linearly on the absorption length of 
$\approx$ 100 m in ice.
 
The expected signal significance in IceCube is somewhat reduced due to two types of correlations between pulses that introduce supra-Poissonian fluctuations. The first correlation involves a single photomultiplier tube. It comes about because a radioactive decay in the pressure sphere can produce a burst of photons lasting several $\mu$s. The second correlation arises from the cosmic-ray muon background; a single cosmic ray shower can produce a bundle of muons which is seen by hundreds of optical modules. 

The 5160 photomultipliers are sufficiently far apart such that the probability to detect light from a single interaction in more than 
one DOM is small. Effectively, each DOM independently monitors several cubic-meters of ice.
The detection principle was demonstrated with the AMANDA experiment, IceCube's predecessor~\citep{bib:AMANDAold}.

The inverse beta process  $\bar \nu_\mathrm{e} + \mathrm{p} \rightarrow \mathrm{e^+} + \mathrm{n}  $  dominates supernova neutrino interactions with $ \mathcal{O}(10\un{MeV})$ energy in ice or water, leading to charged particle tracks of about $0.5 \un{cm} \cdot E_\nu/ \un{MeV}$ length.
Considering the approximate $E_\nu^2$ dependence of the cross section, the light yield per neutrino roughly scales with $E_\nu^3$.
Due to the low rate of galactic supernovae, it is imperative that the detector operates stably for a long time. IceCube was designed to operate for at least 10 years and is well suited for such a purpose owing to an automated online data acquisition, analysis software and alert system. As neutrinos may escape from an exploding supernova at much higher matter densities than photons, neutrinos will be observable several hours before their optical counterpart. The detailed observation of the onset of a supernova explosion is of much interest to astronomers. Since 2009, IceCube has been sending real-time datagrams to the Supernova Early Warning System (SNEWS)~\citep{antonioli-2004-6} when detecting supernova candidate events. SNEWS has been set up to broadcast a reliable alert to the astronomy community when a supernova has been detected by several neutrino detectors within seconds of each other. Currently, Super-Kamiokande~\citep{NuclInstrumMethA.501.418,Ikeda}, LVD~\citep{IlNuovoCimentoA105.1793}, Borexino~\citep{NuclInstrMeth600.568} and IceCube ~\citep{pdd} contribute to SNEWS, with a number of other neutrino and gravitational wave detectors planning to join in the near future.

This paper describes the technical details and expected physics capability of IceCube as a detector for core collapse supernovae.  It also summarizes the performance of the detector while it was still under construction.  
The outline of the paper is as follows: Sect.~\ref{sec:pheno} describes physics processes in supernovae for selected models and oscillation scenarios that are relevant to the performance studies presented in this paper, Sect.~\ref{sec:I3detector} describes the aspects of the IceCube detector relevant to the detection of MeV supernova neutrinos.  In Sect.~\ref{sec:interaction}, we discuss the processes that lead to a detectable signal in IceCube as well as the online analysis that processes and  monitors the data, triggers events and sends out alerts to SNEWS.  Sect.~\ref{sec:detectorperf} describes the performance of the detector over two years and the systematic uncertainties expected when assessing the sensitivity of the detector.  Sect.~\ref{sec:physicsperf} discusses IceCube's potential in the study of astrophysical and neutrino properties, and finally, conclusions are given in Sect.~\ref{sec:conclusions}.

\section{Supernovae and Neutrinos}\label{sec:pheno}
After the core of an aging massive star ceases generating energy and the corresponding radiative pressure from nuclear fusion processes, it undergoes a sudden gravitational collapse as soon as its inactive core grows beyond the Chandrasekhar mass limit. 
After several steps to relieve thermal and degeneracy pressure from the dense electron gas, the collapse stops once nuclear densities are reached and an incompressible proto-neutron star is formed. Matter falling on its surface is promptly stopped and its momentum is inverted forming an outward moving shock wave.

Neutrinos of different flavors are initially trapped in their relative neutrino spheres as the mean free path of neutrinos is smaller than the size of the supernova core at densities larger than \unit[10$^{13}$]{kg/m$^{3}$}. The shock wave following the collapse dissociates nuclei, which suddenly increases the number of protons, resulting in an increase in electron capture and the production of a burst of \nue. The timescale of this neutronization burst (``deleptonization peak'') is on the order of \unit[10]{ms} during which much of the energy driving the shock wave is carried away. The shock stalls but is presumably soon revived by interactions of the large flux of neutrinos generated in the proto-neutron star. The models describing the prompt neutronization burst
appear to be robust and consistent~\citep{PhysRevD.71.063003}. The proto-neutron star subsequently cools over $\sim$ \unit[20]{s}. The neutrino flux decreases until neutrinos are no longer produced in the cooled down proto-neutron star (see e.g. \citet{Fischer} and references therein). 

The released gravitational potential is carried away by huge numbers of neutrinos and to a small extent by heating and expelling the star's outer layers. Less than 1\,\% of the gravitational binding energy of a supernova is emitted as kinetic energy of matter and optically visible radiation. The remaining 99\,\% is released as neutrino energy, of which about 1\,\% will be carried by electron neutrinos from the initial neutronization burst. Most neutrinos and antineutrinos, 
distributed among all flavors, are created during the subsequent cooling processes. An estimated $E_\mathrm{total}\,=\,\int{L_\mathrm{SN}^\mathrm{total}(t')dt'}\approx\,\unit[3\,\times\,10^{53}]{erg}\,=\,\unit[1.87\,\times\,10^{59}]{MeV}$~\citep{PhysRevD.45.3361} is carried away by the intense neutrino burst produced predominantly through thermal Kelvin-Helmholtz cooling reactions~\citep{Suzuki}. Here $L_\mathrm{SN}^\mathrm{total}$ is the time dependent all flavor supernova neutrino and anti-neutrino luminosity. According to~\citet{thompson-2003-592} and ~\citet{PhysRevLett.90.241101}, the mean energy is expected to be about (13 -- 14) MeV for $\nu_\mathrm{e}$, (14 -- 16) MeV for $\bar\nu_\mathrm{e}$ and (20 -- 21) MeV for all other flavors ($\nu_\mathrm{x}$); the Garching model~\citep{AstrAstrop.450.1} differs in that  the mean energies for $\bar\nu_\mathrm{e}$ and $\nu_\mathrm{x}$ turn out to be approximately equal. For a supernova at $d=10$ kpc distance and an average neutrino energy $\overline{E_{\nu}} =$\unit[ 15]{MeV}, the summed flux of all neutrino and antineutrino types, $\Phi_{\rm Earth}^{\rm total}=E_{\rm total}/(4\pi d^2 \overline{E_{\nu}})$, flowing through the detector is $\approx$ \unit[10$^{16}$]{m$^{-2}$}. More on the theory of core collapse supernova can e.g. be found in \citet{janka} and references therein.

In the following paragraphs, we briefly introduce the Lawrence-Livermore~\citep{AstropPhys.496.216} and Garching models~\citep{AstrAstrop.450.1} that are used as benchmarks. In addition, we introduce specific models by~\citet{Dasgupta-2010} and \citet{Sumiyoshi} that were selected to demonstrate IceCube's physics performance in Sect.~\ref{sec:I3performance}. We also discuss the effect of neutrino oscillations on the expected signals, and introduce the parametrization of the energy spectra chosen for this paper. 
  
The spherically symmetric Lawrence-Livermore simulation was performed from the onset of the collapse to \unit[18]{s} after the core bounce, encompassing the complete accretion phase and a large part of the cooling phase. It is modeled after SN 1987A and assumes a 20 $M_\odot$ progenitor. The total emitted energy is $\unit[2.9\,\times\,10^{53}]{erg}$, of which 16\,\% is carried by $\bar\nu_\mathrm{e}$ with \unit[15.3]{MeV} energy on average. 

The newer spherically symmetric Garching simulations include more detailed information on neutrino energy spectra amd use a sophisticated neutrino transport mechanism. They cover \unit[0.80]{s} following the collapse of an O-Ne-Mg \unit[8 -- 10]{} $M_\odot$ progenitor star, that is destabilized due to rapid electron capture on neon and magnesium. This class of stars may represent up to 30\,\% of all core collapse supernovae. Recent simulations by ~\citep{Fischer,Huedepohl} extend over \unit[22]{s} from the collapse of a \unit[8.8]{} $M_\odot$ progenitor to the completed formation of the deleptonized neutron star. They are the only
examples so far where one-dimensional simulations obtain neutrino-powered supernova explosions and two-dimensional simulations yield only minor dynamical and energetic modifcations. In Table~\ref{tab:eventsummary} we also refer to a two-dimensional axisymmetric simulation by~\citet{Marek} of a \unit[15]{} $M_\odot$ progenitor star that covers \unit[0.38]{s} following the collapse. Results with full multi-angle neutrino transport in two dimensions have been reported by~\citet{Brandt}. 

Following original work by \citet{Takahara}, recent simulations \citep{Dasgupta-2010} of certain stellar core-collapse supernovae predict a sharp burst of $\bar{\nu}_\mathrm{e}$ several hundred milliseconds after the prompt $\nu_\mathrm{e}$ neutronization burst associated with a quark-hadron phase transition at high baryon densities. A detection of this prominent feature would constitute direct evidence of quark matter.

The gravitational collapse of less than solar metallicity stars exceeding 25 solar masses will lead to a limited stellar explosion, while stars exceeding 40 solar masses are not expected to explode at all. In both cases a black hole will develop $\mathcal{O}(1\un{s})$ after bounce. At this point, the neutrino emission quickly comes to an end, providing a unique signature for black hole formation~\citep{Sumiyoshi} and a model independent time-of-flight measurement of the neutrino mass~\citep{BeacomBoydMezzacappa}. 

Neutrinos streaming out of the core will encounter matter densities ranging from \unit[$10^{13}$]{kg/m$^3$} to zero. Assuming an energy of 15 MeV, they pass through an MSW-resonance layer at $\approx$\,\unit[$2\times 10^6$]{kg/m$^3$} associated with $\Delta m^2_{23}$ or $\Delta m^2_{13}$, depending on the neutrino mass hierarchy, followed by a second layer at $\approx$\,\unit[$2\times 10^4$]{kg/m$^3$} associated to the quadratic neutrino mass difference $\Delta m^2_{12}$. Both mix the initial fluxes of $\nu_\mathrm{e}$, $\bar\nu_\mathrm{e}$ and $\nu_\mathrm{x}$ depending on the survival probabilities. Although the survival probabilities depend on the details of the density profiles and generic predictions are impossible, we consider three limiting cases as benchmarks to discuss the effect of the assumed neutrino hierarchy on the spectra observed with IceCube. Scenario A describes the normal neutrino hierarchy case and Scenario B represents the inverted hierarchy case with a static density profile of the supernova, both paired with a relatively large mixing angle $\theta_{13} > 0.9^\circ$. In Scenario C, the mixing angle $\theta_{13}$ is assumed to be very small ($\theta_{13} < 0.09^\circ$) and the hierarchy may be either normal or inverted. One should be aware that the predictions are affected by the unknown density profile of the collapsing star. In addition, forwards or backwards running single or multiple shock waves or bubbles can form within the supernova, causing steep density gradients and - in some cases - changes in the oscillation behavior~\citep{Tomas,Choubey}. 

As the neutrinos propagate through the 
Earth, they undergo matter induced oscillations (MSW effect). The 
neutrino flux~\citep{snrate, dighe} decreases by up to 8\,\%. The effect depends sensitively on the zenith angle, the supernova model and the assumed neutrino properties. Given the systematic uncertainties, it will be difficult to establish this effect with IceCube. 
We will therefore include earth oscillation effects in the systematic uncertainty. 

For this paper, the supernova is considered to be close to blackbody source of neutrinos while it is cooling down. The neutrino energies then follow a modified Fermi-Dirac distribution. Many model predictions discussed in this paper adopt the following parametrization for the neutrino differential flux $\Phi_\mathrm{Earth}^\nu$ at the position of the Earth at distance $d$ from the supernova:
\begin{equation}
	\label{flux}
	{ \frac{d\Phi_\mathrm{Earth}^\nu}{dE_{\nu}}= \frac{L_\mathrm{SN}^{\nu}(t)}{4\pi d^2\overline{E_{\nu}}} f(E_{\nu},\overline{E_{\nu}}, \alpha_\nu)}\quad ,
\end{equation}
where
\begin{eqnarray}
	 f(E_{\nu},\overline{E_{\nu}}, \alpha_\nu)  = 
	\left(\frac{1+\alpha_\nu(t)}{\overline{E_{\nu}}(t)}\right)^{1+\alpha_\nu(t)}
	\times  \frac{E_{\nu}^{(\alpha_\nu(t))} \, {\rm e}^{-(1+\alpha_\nu(t)) E_{\nu}/\overline{E_{\nu}}(t)} }{\Gamma(1+\alpha_\nu(t))}
	\label{eq:energydist}
\end{eqnarray}
is the normalized energy distribution depending on a shape parameter $\alpha_\nu$~\citep{keil}. Theory provides the time dependent supernova luminosity 
$L_\mathrm{SN}^\nu(t)$ for the neutrino species $\nu = \nu_\mathrm{e}, \bar\nu_\mathrm{e}, \nu_\mathrm{x}$ with corresponding energies $E_{\nu}$. Other model predictions are transferred to this framework by fitting the provided spectra. 

\section{The IceCube Detector}\label{sec:I3detector}

IceCube~\citep{pdd} was installed in the Antarctic ice sheet at the geographic South Pole between January, 2005 and December, 2010 
by lowering cable assemblies, called strings, into holes drilled in the ice using hot water. IceCube instruments a volume of about \unit[1]{km$^3$} of clear ice between depths of \unit[1450]{m} and \unit[2450]{m} below the surface with a coarse lattice of 5160 Digital Optical Modules (DOMs). Each DOM consists of a photomultiplier tube housed in a pressure-resistant glass sphere. Once the water in the holes refreezes, the DOMs are embedded into the ice sheet with good optical coupling. The DOMs are installed on 86 cable strings, of which 80 will be separated by roughly \unit[125]{m} forming a triangular grid, with each string containing 60 DOMs vertically spaced by \unit[17]{m}. More details on the detector can be found in~\citet{firstyear}. The remaining 6 strings constitute the denser DeepCore sub-array. DeepCore strings are separated by approximately \unit[60]{m} and are located near the center of IceCube. Each of these strings contains 60 high quantum efficiency DOMs, with the bottom 50 DOMs vertically spaced by \unit[7]{m} and located between depths of \unit[(2107 -- 2450)]{m} below a dusty ice layer with reduced transparency. The remaining 10 DOMs, vertically separated by \unit[10]{m} and located above the dust layer, instrument a volume between \unit[(1750 -- 1860)]{m}. 

IceCube is complemented by a surface array called IceTop, consisting of a pair of DOMs encased in ice tanks near the top of each string. Due to their higher noise and sensitivity to the fluctuating solar particle flux~\citep{solar}, the IceTop DOMs do not contribute to the IceCube supernova detection system.   

The DOM is the fundamental element in the IceCube architecture.Each DOM is housed in a 13\inch{} (\unit[33]{cm}) diameter, 0.5\inch{} (\unit[1.27]{cm}) thick borosilicate glass pressure sphere.  It contains a Hamamatsu R7081-02 (R7081-MOD in case of DeepCore) 10\inch{} (\unit[25.4]{cm}) hemispherical photomultiplier tube~\citep{ic3:pmt-paper} as well as several electronics boards containing a processor, memory, flash file system and realtime operating system that allows each DOM to operate as a complete and autonomous data acquisition system~\citep{daq}. It stores the digitized data internally and transmits the information to a surface data acquisition system on request. 

The supernova detection relies on continuous measurements of photomultiplier rates. The rate information  is stored and buffered on each DOM in a 4-bit counter in \unit[1.6384]{ms} time bins ($2^{16}$ cycles of the 40 MHz clock). The main 
data acquisition system ~\citep{daq} transfers the data asynchronously to the independently operating supernova data acquisition system (SNDAQ). For the real-time processing, the information is synchronized by a GPS clock and regrouped in \unit[2]{ms} bins.

The South Pole is out of reach for most communication satellites and high bandwidth connectivity is available only for about 6 hours per day. 
 Therefore, a dedicated Iridium-satellite~\citep{IRIDIUM} connection is used by the SNDAQ host system to transmit urgent alerts. In that case, a short datagram is sent to the northern hemisphere. The receiving system parses the message and forwards information on the supernova candidate event to the international SNEWS group. The time delay between photons hitting the optical module and the arrival of the datagram at SNEWS stands at about \unit[6]{min}, providing close to real-time monitoring and triggering. 
In order to test the signal path, an internal trigger threshold is adjusted to transmit 1 -- 2 background triggers per day.

Due to satellite bandwidth constraints, the data are re-binned in \unit[0.5]{s} intervals and then subjected to a statistical
online analysis described in Sect.~\ref{sec:data_analysis}; the fine time information in \unit[2]{ms} intervals is transmitted for a period starting 30 s before and ending 60 s after a trigger flagging a candidate supernova explosion. The system is surveyed by the IceCube experiment control and monitoring system (``IceCube Live''); supernova alerts are immediately distributed by e-mail to notify experts. 

The optical and noise properties of the DOMs are crucial for the understanding of IceCube's supernova detection and will hence be discussed in more detail in the following subsections. 

\subsection{Optical Properties of the Digital Optical Module}
\label{sec:DOMoptical}
The photomultiplier was chosen on the basis of low dark counts and good time and charge resolution. Its bialkali photocathode has a spectral response in the range \unit[300]{nm} to \unit[600]{nm} with a peak quantum efficiency of $(25 \pm 1)$\,\% at \unit[420]{nm}, well-matched to the Cherenkov signal spectrum and the optical properties of the glacial ice. Dark count rates for standard efficiency DOMs of around \unit[540]{Hz} are typical for DOMs at ice temperatures between \degC{-43} at -1450 m depth and \degC{-20} at -2450 m depth. The quantum efficiency of high efficiency DOMs is roughly 1.35 times higher, while the noise increases approximately by a factor 1.25. The glass of the pressure sphere was selected for high transmission in the sensitive region of the photomultiplier and a low rate of background photons from intrinsic radioactivity in the glass. Optical transparency extends well into the near-UV, with 50\,\% transmission at $T_{50\,\%} \sim \unit[340]{nm}$, but drops to a few percent at \unit[310]{nm}.

\begin{figure}[ht]
\centering
\centering
\includegraphics[angle=0,width=0.45\textwidth]{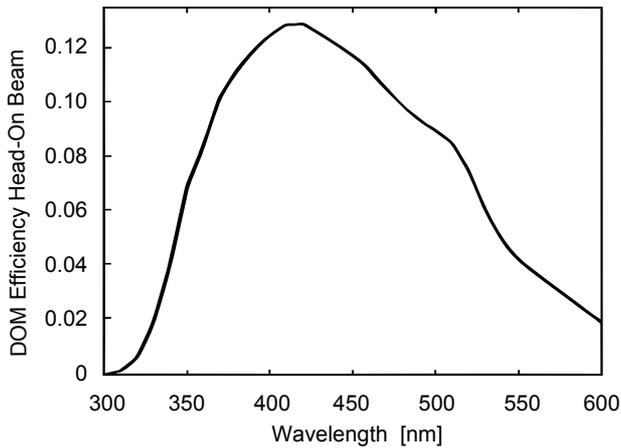}
\caption{
Overall DOM efficiency versus wavelength for head-on illumination of the 0.0856 m$^2$ DOM cross section. The average value in the \unit[300 -- 600]{nm} range, weighted by the wavelength dependence of Cherenkov light emission, is $\approx$ 7.1\%.} 
\label{fig:dom-optical-response}
\end{figure}
The photomultiplier is mechanically attached to the glass pressure housing with a silicone elastomer gel (GE6156 RTV). This gel matches the refractive index of the glass (n=1.48) to reduce optical losses at the medium interfaces. The spectral transparency of the gel extends to approximately \unit[250]{nm} with $T_{50\,\%} \sim \unit[300]{nm}$. The combined response of the glass, gel, and photomultiplier is a critical input to the IceCube Monte Carlo simulation package. The detection efficiency of the DOM to a head-on parallel light beam is shown in 
Fig.~\ref{fig:dom-optical-response}. 

Most of the PMTs are operated at a gain of $10^7$, so single photoelectrons produce pulses of approximately \unit[8]{mV} amplitude and \unit[10]{ns} width across the load impedance. The programmable front end pulse discriminator is set to \unit[2]{mV},
a factor of $\approx 10$ above the RMS noise level~\citep{ic3:pmt-paper}.  On average, 85\,\% of all single photo-electron pulses pass the discriminator threshold.

\subsection{Noise Properties of the Digital Optical Module}
\label{sec:DOMnoise}
\begin{figure}[Ht]
\centering
\includegraphics[angle=0,width=0.45\textwidth]{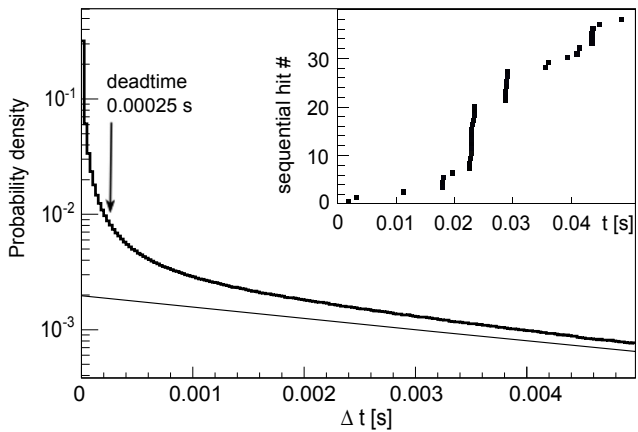}
\caption{Probability density distribution of time differences between pulses for noise (bold line) and the exponential expectation for a Poissonian process fitted  in the range 15 ms $ < \Delta T < $ 50 ms 
(thin line). The excess is due to bursts of correlated hits, as can be seen from the 50 ms long snapshot of hit times shown in the insert.}
\label{fig:dom-noise-response1}
\end{figure}

Several effects contribute to the prevailing average rate of \unit[540]{Hz} for standard efficiency DOMs. Atmospheric muons (\unit[16]{Hz}), thermal emission from the photocathode ($<$\unit[10]{Hz}), and photomultiplier induced afterpulses ($\approx$ \unit[30]{Hz}) all play a role, but the majority of hits are due to radioactive decays of which a large fraction initiates bursts of hits lasting for up to \unit[15]{ms}. These bursts are presumably scintillation of residual cerium in the glass of the photomultiplier and the pressure sphere energized by $\beta$ and $\alpha$ decays. The $^{40}$K content in the IceCube pressure spheres is specified to produce less than \unit[100]{Bq} leaving trace elements from uranium and thorium decay chains as the main source of radioactivity. 

The main characteristics of DOM noise were determined using a minimum bias data set with pulses recorded by 120 IceCube DOMs. The observed time difference between sequential noise hits deviates from an exponential distribution expected for a Poissonian process (see Fig.~\ref{fig:dom-noise-response1}). The inset shows a hit sequence from a single DOM. Photomultiplier related afterpulses, which occur with $\approx$ 6\% probability on time scales of \unit[0.3--11]{$\mu$s}~\citep{ic3:pmt-paper}, cannot explain the high occupancy bursts. 
We infer that the bursts are caused by an event within a DOM, but external to the electron amplification of the photomultiplier.

These observations are consistent with a study by~\citet{HOMeyer2010}, who showed that a photomultiplier with bialkali photocathode produces bursts that increase in rate and size as the photomultiplier is chilled. Such behavior could result from increased efficiency for radiative decay of excited states in the glass. In addition, Richardson's law describes the increase in thermal emission with photomultiplier temperature. The deployment of IceCube DOMs on vertical strings places them in environments ranging 
from \degC{-43} to \degC{-20}~\citep{PRICE}, warming with depth. DOM temperatures are some \degC{10} warmer due to the energy dissipated in the electronics as monitored by a sensor mounted on the mainboard, but the photomultiplier and DOM glass temperature is somewhat uncertain. 

\begin{figure}[Ht]
\centering
\includegraphics[angle=0,width=0.45\textwidth,bb= 0 0 515 385]{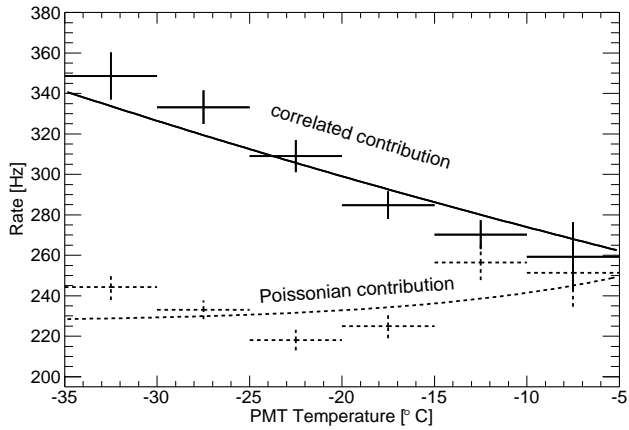}
\caption{Measured average noise rates of 120 DOMs as function of DOM temperature. The excess of short time intervals due to bursts  (solid) is fitted to the empirical model of~\citet{HOMeyer2010}. The Poissonian contribution (dashed) has not been corrected for the depth (and thus temperature) dependent contribution of atmospheric muons (see Fig.~\ref{fig:long_term_rates}). The dotted line is a comparison to the predicted thermal noise from Richardson's law plus a constant rate of $C= 225 \pm  6$ Hz. }
\label{fig:dom-noise-response2}
\end{figure}

To confirm these temperature patterns, we divide the DOM noise into two contributions. The first is the rate of random arrivals as determined by fitting the slope of the interval distribution as in Fig.~\ref{fig:dom-noise-response1}. The second is the rate of events contributing to the excess of short intervals. These contributions are further divided into six temperature bands, and displayed in Fig.~\ref{fig:dom-noise-response2}. The fitted excess contributions are then compared to an empirical exponential ansatz~\citep{HOMeyer2010},
\begin{equation}
	r(T) = G \cdot A_C \cdot e^{-\frac{T}{T_r}} \quad ,
\end{equation}
where $T$ is the absolute temperature, $A_C=0.055$ m$^2$ is the cathode surface of the deployed R7081-02 photomultiplier \citep{hamamatsu} and $G=5\times 10^4$ /m $^2$/s is a fixed constant taken from ~\citet{HOMeyer2010}.  The fit results in $T_{r}\approx 115$ K. 
The Poisson component is fitted to 
the Richardson-type law on thermal emission plus an ad-hoc constant noise term $C$
\begin{equation}
	r(T) = A \cdot T^{2} \cdot e^{-\frac{W}{kT}}  + C \quad ,
\end{equation}
where $k$ denotes Bolzmann's constant, $W\approx 0.5 $ eV is the work function for Bialkali cathodes, and $A$, $C$ are fit parameters.
The curves match the observed temperature dependence of the noise rates fairly well, and support the hypothesis that the DOM noise is primarily due to a temperature dependent spectrum of bursts. 

The data acquisition was designed to reduce the noise rate by eliminating the excess hits, while keeping the random arrivals. 
The signal-to-noise ratio of the measurement can be improved by enforcing an artificial dead time $\tau$ after every count, configured to \unit[250]{$\mu$s} by a field programmable gate array in the DOM.  This reduces the noise rate from \unit[540]{Hz} to \unit[286]{Hz} at the cost of some 13\% dead time for signal. The choice of \unit[250]{$\mu$s} optimizes sensitivity to the Lawrence-Livermore model~\citep{AstropPhys.496.216} for distances up to \unit[75]{kpc}, when neglecting the effect of afterpulses following the signal. A dead time $\tau > 110$ $\mu$s guarantees that the 4-bit counters do not overflow. The rate decrease due to dead time can be corrected for, however, the corresponding uncertainty increases once the measured rate approaches 1/$\tau$. An improved data acquistion that would avoid distortions of the rate measurement for supernova distances $< \mathcal{O}(1\un{kpc})$ is under discussion. 
Various dead time implementations, e.g. schemes masking hits that arrive within the dead time caused by a previous hit or schemes allowing each hit to restart the dead time, were tested with only slight differences observed. By eliminating the initial hits of the bursts, the noise rate can be reduced by up to 100 Hz. Other optimizations may be possible but require a thorough understanding of the effect on signal hits. Design and implementation of a new data acquisition system with more efficient noise rejection will be reported in a subsequent publication.

\section{Neutrino Interaction, Detection and Analysis}\label{sec:interaction}
In this section we first discuss the processes that lead to a detectable signal in IceCube, starting with the relevant neutrino interaction cross sections and continuing with the Cherenkov light production, propagation and detection. The real-time analysis introduced to monitor collective rate changes in IceCube's light sensors is described in the second part of this section.  
 
\subsection{Cherenkov Photon Signal in IceCube}
\label{sec:I3signal}

\begin{table*}[t]
\caption{\label{tab:cross} Major neutrino reactions.}
\centering
\begin{tabular}{llccc}
\hline\hline

Reaction & $\#$ Targets & $\#$ Signal Hits & Signal Fraction  & Reference \\

\hline

		$\bar\nu_\mathrm{e} + \mathrm{p} \rightarrow \mathrm{e^+ + n}$ & $6\cdot 10^{37}$ & 134 k (157 k) & 93.8\,\% (94.4\,\%) & \citet{strumia}\\
		$\nu_\mathrm{e} + \mathrm{e}^-\rightarrow \nu_\mathrm{e} + \mathrm{e^-}$ & $3\cdot 10^{38}$ & 2.35 k (2.25 k) & 1.7\,\% (1.4\,\%) &  \citet{marciano:03}\\
		$\bar\nu_\mathrm{e} + \mathrm{e^-} \rightarrow \bar\nu_\mathrm{e} + \mathrm{e^-}$ & $3\cdot 10^{38}$ &660 (720)  & 0.5\,\% (0.4\,\%)  & \citet{marciano:03}\\
		$\nu_\mathrm{\mu+\tau} + \mathrm{e^-} \rightarrow \nu_\mathrm{\mu+\tau} + \mathrm{e^-}$ & $3\cdot 10^{38}$& 700 (720) & 0.5\,\% (0.4\,\%)  & \citet{marciano:03}\\
		$\bar\nu_\mathrm{\mu+\tau} + \mathrm{e^-} \rightarrow \bar\nu_\mathrm{\nu+\tau} + \mathrm{e^-}$ & $3\cdot 10^{38}$ & 600 (570) & 0.4\,\% (0.4\,\%)  & \citet{marciano:03}\\
$\nu_\mathrm{e} + {^{16}}\mathrm{O} \rightarrow \mathrm{e^-} + {\rm X}$ & $3\cdot 10^{37}$ & 2.15 k (1.50 k) & 1.5\,\% (0.9\,\%)  & \citet{Kolbe}\\
$\bar\nu_\mathrm{e} + {^{16}}\mathrm{O} \rightarrow \mathrm{e^+} + {\rm X}$& $3\cdot 10^{37}$ & 1.90 k (2.80 k) & 1.3\,\% (1.7\,\%) & \citet{Kolbe}\\
$\nu_\mathrm{all} + {^{16}}\mathrm{O} \rightarrow \nu_\mathrm{all} + {\rm X}$& $3\cdot 10^{37}$ & 430 (410) & 0.3\,\% (0.3\,\%) &  \citet{Kolbe}\\
$\nu_\mathrm{e} + {^{17/18}}\mathrm{O}/^2_1\mathrm{H}\rightarrow \mathrm{e^-} + {\rm X}$ & $6\cdot 10^{34}$ & 270 (245) & 0.2\,\% (0.2\,\%) &  \citet{HaxtonR}\\
\hline
\end{tabular}
\tablefoot{The approximate number of targets in a 1 km$^3$ ice detector, the detected number of hits at 10 kpc distance and their fraction  in stars are given in the second, third and fourth column, respectively. In order to indicate the effect of neutrino oscillations in the star, signal hits and fractions are presented both assuming a normal neutrino hierarchy (Scenario A) and - in brackets - assuming an inverted hierarchy (Scenario B). The numbers are taken from the Garching model using the equation of state by~\citet{Lattimer}, integrating over \unit[0.8]{s} and averaging over the neutrino incidence angle. }
\end{table*}

Neutrinos of different species will be detected in IceCube via the interactions listed in Table~\ref{tab:cross}. The table also includes the number of observed photon hits and the corresponding fractions as expected from the Garching model. The inverse beta reaction dominates in the ice. The small contribution by neutrino-electron scattering processes poses a challenge to the detection of the deleptonization peak. Note  
that the $\nu_\mathrm{e}$ and $\bar{\nu}_\mathrm{e}$ cross sections on $^{16}$O are strongly energy dependent due to the high reaction thresholds of 15.4 MeV and 11.4 MeV, respectively. While their contribution to the hit rate in case of the Garching model is small ($\approx 3$\,\%), the contribution can be as high as 20\,\% for a 40 $M_\odot$ progenitor~\citep{Sumiyoshi} with average neutrino energies of 25 MeV and beyond.  The $\nu_{\rm e}$ cross sections on  the rare isotopes $^{18}{\rm O}$,  $^{17}{\rm O}$ and on $^2_1{\rm H}$, with $^{18}{\rm O}$ giving the dominant contribution, add a small signal due to their low reaction thresholds~\citep{Haxton}.
As the cross sections are only given for electron energies between (5--13) MeV~\citep{HaxtonR}, they were extrapolated assuming an $E^2_\nu$ dependence. While the energy deposition due to positron annihilation, neutron capture and photon induced compton electrons arising in the de-excitation of giant O* resonances in the neutral current interactions $\nu_{\rm X} + ^{16}{\rm O}\rightarrow \nu_{\rm X} + ^{16}{\rm O^*} \rightarrow \nu_{\rm X} + ^{15}{\rm O}/^{15}{\rm N} + {\rm n/p} + \gamma$ ~\citep{Langanke} have been included, we have not yet considered delayed $\beta$ and $\beta\gamma$ decays from excited nuclei. The cross section uncertainties of the reactions on protons and electrons are estimated in the references of Table~\ref{tab:cross} to be smaller than 1\,\%; uncertainties on oxygen reactions are hard to assess and the cross section may be only known up to a factor of two.

\begin{figure}[ht]
\centering
\includegraphics[angle=0,width=0.45\textwidth,bb=0 0 524 436]{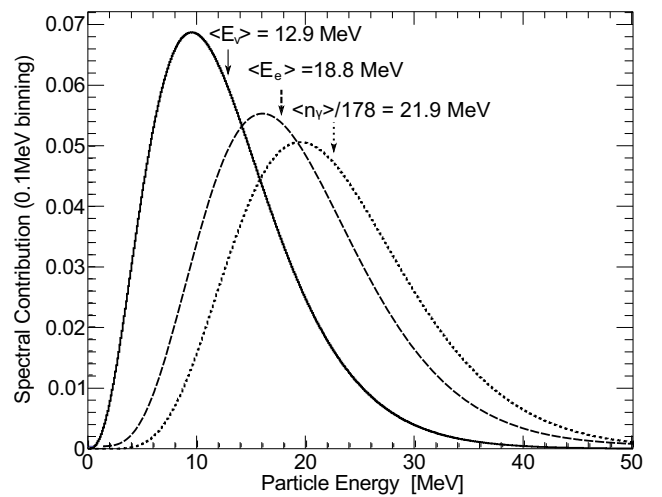}
\caption{Example for an anti-electron neutrino energy spectrum with $\alpha=2.92$ predicted for an 8.8 solar mass O-Ne-Mg core supernovae collapse~\citep{Huedepohl} one second after the onset of the burst (solid line). Also shown is the cross section weighted energy spectrum of produced positrons and electrons from inverse beta decay (dashed line) as well as a measure of the detectable energy, which is proportional to the number of Cherenkov photons $N_\gamma$ in the (300 -- 600) nm range (dotted line). 
The relation $ N_{\gamma} = 
a \cdot E_\mathrm{e}$ with $a= 
178$ MeV$^{-1}$ is used.}
\label{energy_spectrum}
\end{figure}

Reactions producing electrons or positrons in the final state radiate $N_{\gamma}$ Cherenkov photons along their flight path $x$, as long as their kinetic energies exceed the Cherenkov threshold of 0.272 MeV. Integrating the Frank-Tamm formula between (300 -- 600) nm and accounting for the dispersion in ice, one obtains  $N_{\gamma}=(316 \pm 9)$ cm$^{-1}\cdot x$ ($x$ in cm). 
For the inverse beta decay, the total average positron energy $\overline{E_\mathrm{e^+}}$ is calculated from the average $\bar\nu_\mathrm{e}$ energy  by the relation $\overline{ E_\mathrm{e^+}} \approx
(\overline{E_{\bar\nu_\mathrm{e}}}-\delta)\cdot \left[1-\overline{E_{\bar\nu_\mathrm{e}}}/(\overline{E_{\bar\nu_\mathrm{e}}}+m_\mathrm{p}c^2)\right]$ with 
$\delta=(m_\mathrm{n}^2-m_\mathrm{p}^2-m_\mathrm{e}^2)/2m_\mathrm{p}$. Due to the approximately quadratic energy dependence of the interaction cross section, the observed positron energies are on average higher than those of the incoming neutrino; the number of Cherenkov photons approximately rises with $E_\nu^3$ (see Fig.~\ref{energy_spectrum}).

The mean travel path for $\mathcal{O}(\unit[10]{MeV})$ electrons from $\nu_\mathrm{e}$ and positrons from $\bar\nu_\mathrm{e}$, including secondary leptons with energies above the Cherenkov threshold as well as  positron annihilation, was determined with a GEANT-4 Monte Carlo simulation. The linear relationship $\bar{x} = (0.560 \pm 0.005 \,\rm{(stat.)}\pm 0.030 \,\rm{(syst.)})\un{cm}$ $\cdot E_{\mathrm{e}^+}/\un{MeV}$ was found for positrons. The corresponding relationship for electrons was determined to be consistent within errors.

The optical scattering and absorption in glacial ice at the South Pole has been studied extensively~\citep{iceproperties06} by the AMANDA and IceCube Collaboration with pulsed and continuous light sources embedded in the ice with the neutrino telescope. The detectors span depths ranging from \unit[(1300 -- 2500)]{m} in the ice where the scattering coefficient varies by a factor of seven and absorptivity can vary by a factor of three depending on the wavelength. The data~\citep{Bramall} are consistent with the variations in dust impurity concentration seen in ice cores sampled at other Antarctic sites to track climatological changes. In the simulation applied for this paper, the ice is assumed to be homogeneous in the horizontal plane despite an observed slight tilt. 

We use two alternative procedures to calculate the number of detected signal hits from the number of neutrinos crossing the detector: the first approach is based on separate simulations of particle interactions, Cherenkov photon creation, propagation and detection, the second  GEANT-3.21 GCALOR-based~\citep{geantgcalor} simulation combines all the steps in one program. 

\looseness-1 
IceCube's standard simulation of photon propagation within the ice relies on predetermined tables \citep{Lundberg:2007yg}, created to track photons across the Antarctic ice.
The tables store the detection probability and the arrival time distribution for given source and detector locations as well as their orientation. It includes the source wavelength, angular and intensity information, DOM parameters such as the glass and gel transparency and the quantum efficiency of the photomultiplier tubes. It also contains information about the ice such as the depth-dependent absorption and scattering lengths. 

The signal hit rate per DOM for a specific reaction and target is given by:
\begin{eqnarray}
	R(t) & = & \epsilon_\mathrm{dead time}\frac{n_\mathrm{target} \,L_\mathrm{SN}^\nu(t)}{4\pi d^2\overline{E_{\nu}}(t)} 
	 \int_0^{\infty} \,dE_\mathrm{e} \!\! \int_0^{\infty}  \,dE_{\nu} \nonumber \\
	&\times\, & \frac{d\sigma}{dE_\mathrm{e}}(E_\mathrm{e},E_{\nu}) \, N_{\gamma}(E_\mathrm{e}) \,V_\mathrm{\gamma}^\mathrm{eff} \,f(E_{\nu},\overline{E_{\nu}}, \alpha_\nu,t) \, , 
	  \label{eq:rate}
\end{eqnarray}
where $n_\mathrm{target}$ is the density of targets in ice, $d$ is the distance of the supernova, $L_\mathrm{SN}^\nu(t)$
its luminosity, $f(E_{\nu},\overline{E_{\nu}}, \alpha_\nu, t)$ is the normalized neutrino energy distribution defined in Eq.~\ref{eq:energydist} and  $E_\mathrm{e}$ denotes the energy of electrons or positrons emerging from the neutrino reaction. $V^\mathrm{eff}_\gamma$ denotes the effective volume for a single photon and $N_\gamma(E)\approx 178\cdot E_\mathrm{e}$ is the energy dependent number of radiated Cherenkov photons; their numerical values depend on the selected wavelength range, chosen as (300 -- 600) nm throughout this paper. 
The artificial dead time  $\tau$ (see Sect.~\ref{sec:DOMnoise}) reduces the total rate of hits. Comparing the observed signal, defined as the net increase over the nominal noise level, to the full rate of signal hits defines the dead time efficiency $\epsilon_\mathrm{dead time}$.
The approximate expression $\epsilon_\mathrm{dead time}\approx 0.87/(1+r_{SN}\cdot \tau)$ is found as function of signal rate $r_{SN}$ by adding Poissonian signal to the measured sequence of noise hits and applying a non-paralyzable dead time  $\tau=250$ $\mu$s. 

The single photon effective volume varies strongly with the photon absorption. As a first approximation, $V_\gamma^\mathrm{eff}$ can be estimated by the product of the Cherenkov spectrum and DOM sensitivity weighted absorption length ($\approx \,$\unit[100]{m}), DOM geometric cross section (\unit[0.0856]{m$^2$}), Cherenkov spectrum weighted optical module sensitivity ($\approx \,$0.071), average angular sensitivity including cable shadowing effects ($\approx\,$0.32), and the fraction of single photon hits passing the electronic DOM threshold ($\approx \,$0.85). 

$V_\gamma^\mathrm{eff}$ was simulated by randomly placing $10^7$ photons with (300 -- 600) nm wavelengths within a sphere of radius \unit[250]{m} around each DOM. We made the simplifying assumption that the Cherenkov light arrives at the DOMs isotropically from all angles. Note that the directions of positrons from the dominant inverse beta decay reaction are very weakly correlated with those of the incoming neutrinos. 

\begin{figure}[ht]
\centering
\includegraphics[angle=0,width=0.48\textwidth]{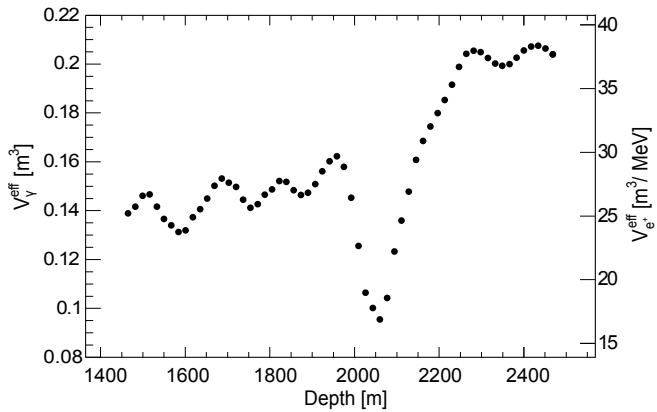}
\caption{Effective volume $V_{\gamma}^\mathrm{eff}$ per DOM (left axis) for detection of 
 Cherenkov photons with (300 - 600) nm wavelength plotted as a function of depth. The effective positron volume can be read off the right axis. DeepCore strings are not included in these plots. }
\label{veffplot}
\end{figure}

$V_{\gamma}^\mathrm{eff}$ was determined as function of depth in the ice (see Fig.~\ref{veffplot}). Averaging over all DOMs in one string
one obtains   $V_{\gamma}^\mathrm{eff} = 0.163 \pm 0.004 \,\rm{(stat.)} \pm 0.020 \,\rm{ (syst.)} \un{m^3}$. The systematic uncertainty
 is discussed in Sect.~\ref{sec:systematics}.

The energy dependent effective volume $V_{\mathrm{e}}^\mathrm{eff}$ for detecting an electron or positron is obtained by multiplying $V_{\gamma}^\mathrm{eff}$ with the number $N_\gamma(E)$ of Cherenkov photons.
The mean number of photons recorded by an optical module averaged over energy is then given by 
$N^\mathrm{detect}_\gamma = \epsilon_\mathrm{dead time}\cdot n^{\mathrm{interact}}_\nu\,\cdot\,\overline{V^\mathrm{eff}_\mathrm{e}}$, 
where $n^{\rm interact}_\nu$ is the neutrino density. For positrons with a cross section weighted average energy of e.g. $\overline{E_\mathrm{e^+}}$=20 MeV (see Fig.~\ref{energy_spectrum}) one would obtain the average effective volume $\overline{V_\mathrm{e}^\mathrm{eff}}= (29.0 \pm 3.8)\, \mathrm{m^3/MeV}\, \cdot \, \overline{E_\mathrm{e^+}} \approx (580\pm 80)\, \mathrm{m^3}$ for standard efficiency DOMs. This volume corresponds to an envisioned sphere of $\approx 5.2\un{m}$ radius centered at the optical module position, with full sensitivity inside and zero outside. With 5160 optical modules deployed, IceCube thus roughly corresponds to a dedicated 3.5 Mton supernova search detector in terms of geometry. Due to the presence of noise, a fair comparison in terms of statistical accuracy needs to take into account the signal over background ratio as function of time and distance. To give an example, a study of the initial 380 ms of the burst in the Lawrence Livermore model (see Table~\ref{tab:eventsummary}) at distances of 10 kpc (5 kpc) would require a 0.45 (1.6) Mton background free detector to statistically compete with IceCube.

\begin{figure}[ht]
\centering
\includegraphics[angle=0,width=0.42\textwidth]{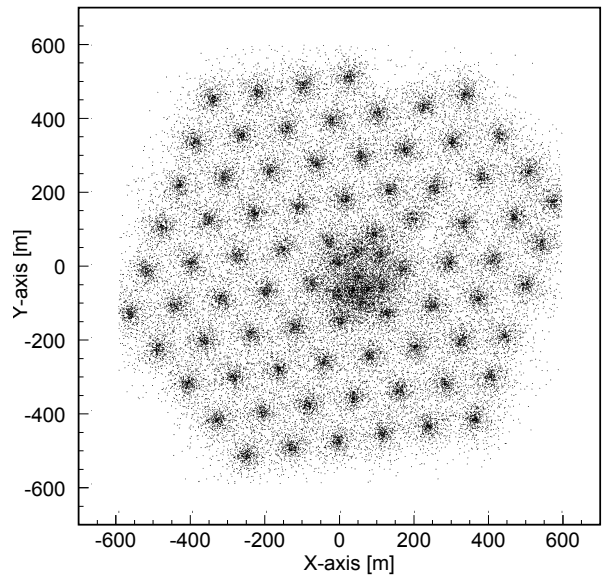}
\caption{Detected neutrino inverse beta decay interaction vertices projected onto the horizontal plane based on a GEANT-3.21 simulation with 10 million neutrino interactions. }
\label{xyplot}
\end{figure}

\looseness-1
The second approach was to apply a GEANT-3.21 GCALOR based simulation of individual events that includes $\nu_\mathrm{e}$ and $\bar{\nu}_\mathrm{e}$ on protons, electrons and $^{16/17/18}\mathrm{O}$, positron annihilation and neutron capture, the photon propagation in the ice including the effect of dust layers, detector geometry, and the DOM response~\citep{Richardthesis}. The $\bar{\nu}_\mathrm{e}+\mathrm{p}\rightarrow \mathrm{e^+ + n}$ cross section parametrization of~\citet{PhysRevD.60.053003}, which is in good agreement with~\citet{strumia}, was used. Positron annihilation and hydrogen capture of neutrons produce
photons of 0.51 MeV and 2.22 MeV energy, respectively. These add, predominantly by Compton scattering and subsequent Cherenkov emission, $\approx$ 1 MeV to the recorded energy. 
Rates from neutrino interactions on electrons reveal a 20\,\% dependence on the incoming neutrino direction due to the small angle between neutrinos and scattered electrons and a directional dependence of the DOM efficiencies. 
Fig.~\ref{xyplot} shows the clustering of detected inverse beta neutrino interactions at the position of the detector strings to visualize the effective volumes.
The use of events with two or more DOMs detecting photons from the same positron to improve upon IceCube's sensitivity at large supernova distances and to track relative changes in the average neutrino energy will be discussed in a future paper. If several photons arrive close in time at the same DOM they will be counted as one hit; if one of the photons is delayed by scattering it will be rejected by the artificial dead time requirement. 
The two independent approaches for the determination of the detected number of events agree within 10\,\%.

One may obtain a rate estimate from measured data by scaling
the 11 events Kamiokande-II observed during the Supernova SN1987A neutrino burst to IceCube's effective volume of Antarctic ice. Assuming the $\nu$ energy spectrum of~\citet{bib:Vissani}, accounting for the Kamiokande-II energy threshold and positron detection efficiency, and taking into account the loss due to IceCube's artificial dead time
we determine a signal expectation of 
$113 \pm 36$ 
detected photons per IceCube module within the first \unit[15]{s} for a SN1987A like supernova near the galactic center at \unit[10]{kpc} distance. 
The results are consistent with earlier simulations~\citep{FeserThesis, jacobsen} performed for AMANDA that assumed homogeneous ice, after correcting for the different photomultiplier sensitive areas, optical module transparencies and dust layers in the ice.

\subsection{Real-Time Analysis Method}
\label{sec:data_analysis}

The analysis monitors the collective rate increase $\Delta\mu$ in all DOMs induced by Cherenkov photons uniformly distributed in the ice. As discussed in Sect.~\ref{sec:I3signal}, the photons are radiated by $\mathrm{e}^\pm$ produced by reacting supernova neutrinos. Counting $N_i$ pulses during a given time interval $\Delta t$, rates $r_i=N_i/\Delta t$ 
for DOM $i$, are derived.  The index $i$ ranges from $1$ to the total number of operational optical modules $N_\mathrm{DOM}$. With sufficiently large $\Delta t$'s, the distributions of the $r_i$'s can be described by lognormal distributions that, for simplicity, are approximated by Gaussian distributions with rate expectation values 
$\langle r_i\rangle$ and corresponding standard deviation expectation values $\langle\sigma_i\rangle$.
These 
expectation values are computed from moving \unit[300]{s} time intervals before and after the investigated time interval. Shorter time intervals reduce the sensitivity of the analysis. At the beginning and the end of a SNDAQ-run, asymmetric intervals are used.  The time windows exclude $30\un{s}$ before and after the investigated bin in order to reduce the impact of a wide signal on the mean rates. 

 The most likely \textit{collective rate deviation} $\Delta\mu$ of all DOM noise rates $r_i$ from their individual $\langle r_i\rangle$'s, assuming the null hypothesis of no signal, is obtained by maximizing the likelihood
\begin{equation}
	\mathcal{L}(\Delta\mu) = \prod_{i=1}^{N_\mathrm{DOM}} \, \frac{1}{\sqrt{2\pi}\,\langle\sigma_i\rangle} \, {\rm exp}(-\frac{(r_i-(\langle r_i\rangle+\epsilon_i\,\Delta\mu))^2}{2\langle\sigma_i\rangle^2}) \,\, .
\end{equation}
Here $\epsilon_i$ denotes a correction for module and depth dependent detection probabilities. An analytic minimization of $-\ln\mathcal{L}$ leads to
\begin{equation}
	\Delta\mu = \sigma_{\Delta\mu}^2 \sum_{i=1}^{N_\mathrm{DOM}} \, \frac{\epsilon_i\,(r_i - \langle r_i\rangle)}{\langle\sigma_i\rangle^2} \quad ,
\end{equation}
with an approximate uncertainty of
\begin{equation}
	\sigma_{\Delta\mu}^2 = \left(\sum_{i=1}^{N_\mathrm{DOM}} \, \frac{{\epsilon_i}^2}{\langle\sigma_i\rangle^2}\right)^{-1} \quad .
\end{equation}
Note that $\Delta\mu$ has the structure of a weighted average sum: each optical module contributes with the deviation of its expected noise rate weighed by 
$\epsilon_i/\langle\sigma_i\rangle^2$. 
Assuming uncorrelated background noise and a large number of contributing DOMs, the significance $\xi = \Delta\mu/\sigma_{\Delta\mu}$ should approximately follow a Gaussian distribution with unit width centered at zero.  In practice, the width turns out to be larger (see Sect.~\ref{sec:significance}). The likelihood that a deviation is caused by an isotropic and homogeneous illumination of the ice can be calculated from the $\chi^2$-probability
\begin{equation}
	-2\ln({\mathcal L})=\chi_{\Delta\mu}^2 = \sum_{i=1}^{N_\mathrm{DOM}} \left(\frac{r_i-(\langle r_i\rangle+\epsilon_i\,\Delta\mu)}{\langle\sigma_i\rangle}\right)^2.
\end{equation}
In order to suppress high rate deviations due to a temporary malfunction of individual detector modules, we reject supernova candidate events with a
$\chi_{\Delta\mu}^2$-probability $< 0.001$. 

For short time bases $\Delta t$, the Gaussian approximation is no longer valid and Poissonian probabilities must be used. The collective rate deviation can then be obtained from the equation
\begin{equation}
	0 = \frac{d\,\ln({\mathcal L}(\Delta\mu))}{d\,\Delta\mu}= \sum_{i=1}^{N_\mathrm{DOM}} \Bigg\{ \frac{n_i \epsilon_i}{\langle r_i\rangle + \epsilon_i \Delta\mu} - \epsilon_i\Bigg\}\quad ,
\end{equation}
which can no longer  be solved analytically. The same goes for the corresponding uncertainty, which is derived by identifying a drop in $\ln{\mathcal L}$ by $0.5$. The required numerical minimization prevents an online analysis of the raw data in 2 ms time intervals. However, fine time data in intervals of 30 s before and 60 s after a trigger are transmitted by satellite to perform a more detailed analysis offline. 
For instance, the onset of the neutrino emission can be determined with better than \unit[5]{ms} accuracy for  
supernovae with less than 15 kpc distance conservatively assuming low mass O-Ne-Mg core supernovae~\citep{Huedepohl}. For similar studies see~\citet{Pagliaroli-a} and~\citet{halzen+raffelt}. This information can then be used to triangulate the supernova direction with other neutrino experiments.

The optimal time base $\Delta t$ to detect faint signals depends on the expected signal shape. A simple generic description incorporates a fast $\mathcal{O}(10\un{ms})$ rise of the neutrino flux followed by an exponential decrease as expected during proto neutron star cooling with a time constant of $\tau \approx \unit[3]{s}$.
Maximizing
\begin{equation}
	\frac{\Delta\mu}{\sigma_{\Delta\mu}} \propto \frac{\int_0^{\Delta t}\,e^{-\frac{t}{\tau}}\,dt}{\sqrt{\Delta t}}
\end{equation}
leads to $\Delta t\approx 1.26\, \tau \approx \,\unit[3.8]{s}$. As the realtime analysis operates on bins of \unit[0.5]{s} length, a time window length of \unit[4]{s} has therefore been chosen as the best available setting for this particular model assumption. Assuming the Livermore model~\citep{AstropPhys.496.216}, with a pronounced flux during the first seconds due to the high mass progenitor, the optimal time window is determined to be \unit[1.6]{s}.

To cover these model uncertainties, additional analyses with time bases of \unit[0.5]{s} and \unit[10]{s} are run in parallel with the one with $\Delta t$ = \unit[4]{s}. The \unit[0.5]{s} analysis aims at
short neutrino bursts (i.e. from soft gamma ray repeater sources or from
supernovae collapsing into a black hole). The \unit[10]{s} time base accounts for the observed time window of the detection of the neutrinos from the supernova SN1987A by Kamiokande-II~\citep{PhysRevLett.58.1490}.
By removing the cut on $\chi^2$ for the \unit[0.5]{s} binning, the trigger has been made sensitive to partial illuminations of the detector. This gives the possibility to record hypothetical exotic particles emitting considerable amounts of light and thereby acting as a slowly moving source (such as ultra-heavy magnetic monopoles in some theories).

The collective rate deviation $\Delta\mu$ and its uncertainty $\sigma_{\Delta\mu}$ in the time bases of \unit[4]{s} and \unit[10]{s} are calculated using sliding windows in \unit[0.5]{s} steps and extracting the maximal significance. This procedure ensures that the signal detection efficiency is not reduced by binning effects. 

\looseness-1
\section{Detector Performance}\label{sec:detectorperf}

In this section we will characterize the detector performance based on two years of data taking experience, discuss detector qualification criteria and summarize the expected systematic uncertainties. The data were taken with 22 operating strings (211 days between Aug 2, 2007 and April 5, 2008) and with 40 operating strings (345 days between April 9, 2008 and April 15, 2009).
 
\subsection{DOM Stability Requirements}
\label{sec:qualification}
\looseness-1
The stability of DOM noise rates is crucial for IceCube's sensitivity to detect supernovae. Faulty modules are removed from the analysis using automatic procedures that are applied in real time. 
In the 40 string configuration, 41 DOMs out of 2400 deployed DOMs showed no signal ($\approx$ 1.7\,\%); all module rates fulfilled the requirement $\langle r_i\rangle < 10000\un{Hz}$. 
Operational modules are removed from the analysis if they exhibit a variance $\langle \sigma_i \rangle^2$ much larger than the Poissonian expectation $\langle r_i\rangle$  or high skewness $|s|$. 
In the very rare case, where the number of qualified modules drops below a threshold of 100, the corresponding time periods are discarded as a safeguard to prevent sending false alarms to SNEWS.

The filter results in Table~\ref{tab:filtered} show the excellent data quality of IceCube. 
Taking the 40 string configuration as an example,
98\,\% of the disqualifications were due to just 11 DOMs. Thus the module disqualification has a negligible effect on the signal significance which changes as the square root of active DOMs. 

\begin{table*}[ht]
\caption{\label{tab:filtered} Module disqualification.} 
\centering
\begin{tabular}{lccccc}
\hline\hline

Analysis       & Broadening cut & Loss (\%)  & Skewness cut & Loss (\%) & Fraction of DOMs removed \\
bin width (s)  &                &            &              &           & from analysis (\%)       \\

\hline
		\unit[0.5]{}  & 0.64  $< \langle\sigma_i\rangle^2/\langle r_i\rangle<$4.0  & 0.02 & $|s| < $ 0.8   & 0.01  & 1.73 \\
		\unit[4]{}    & 0.64  $< \langle\sigma_i\rangle^2/\langle r_i\rangle<$4.0  & 0.05 & $|s| < $ 1.2   & 0.02  & 1.75 \\
  	\unit[10]{}   & 0.36  $< \langle\sigma_i\rangle^2/\langle r_i\rangle<$6.25 & 0.02 & $|s| < $ 1.7   & 0.01  & 1.73 \\
\hline
\end{tabular}
\tablefoot{Disqualification requirements used in the online analysis and corresponding average percentage of DOMs that are rejected.  The values were extracted from successful data taking runs with 40 IceCube strings covering the time from 2008/04/09 to 2009/04/19, which corresponds to an uptime of about 345 days. 
Most DOMs removed from the analysis are dysfunctional and provide zero rate.}
\end{table*}

\looseness-1
\subsection{Long Term Stability}
\label{sec:long_term}
\begin{figure}[ht]
\centering
\includegraphics[angle=0,width=0.46\textwidth]{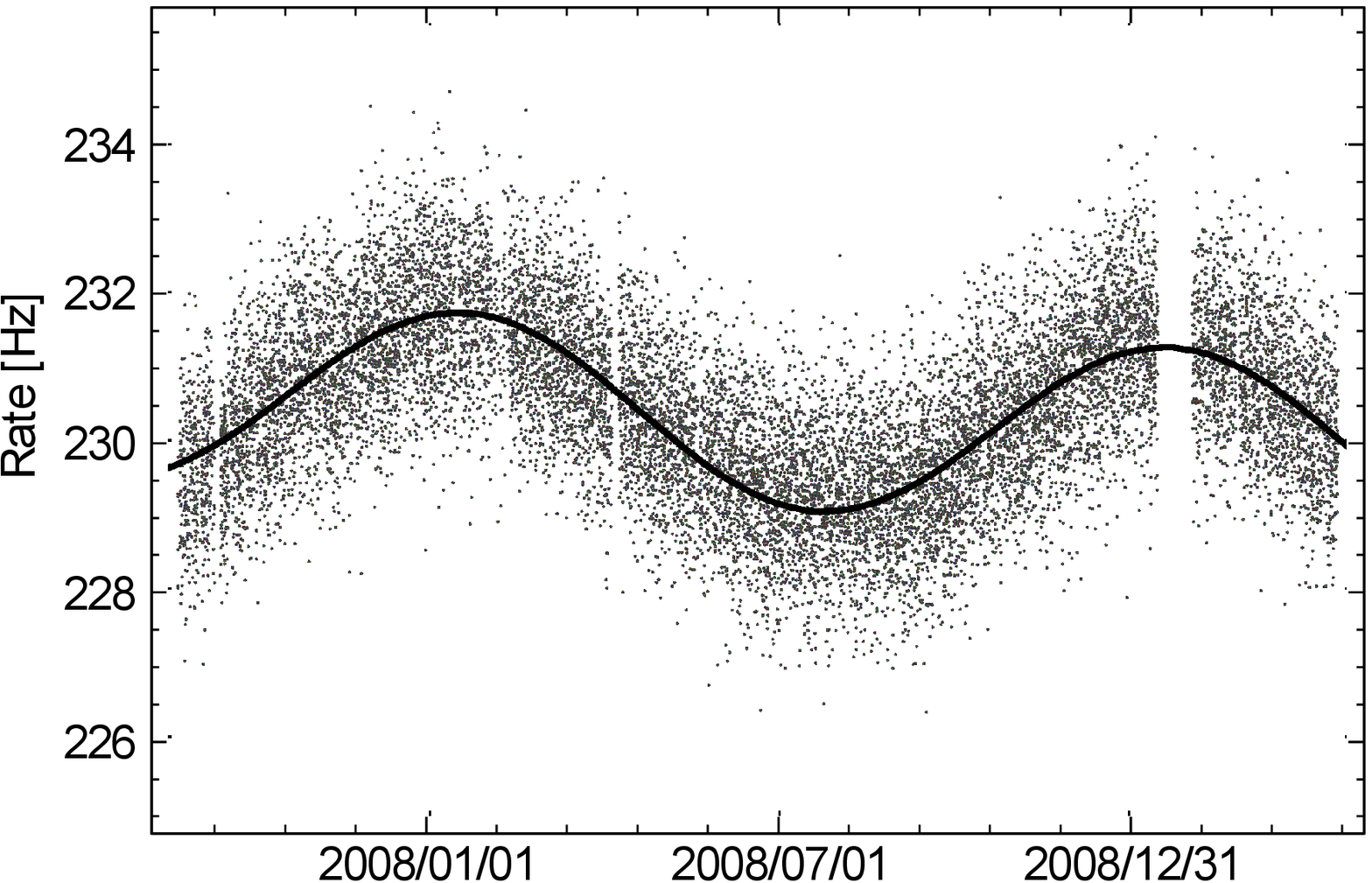}\vspace{20pt}
\includegraphics[angle=0,width=0.46\textwidth,bb=0 -20 569 389]{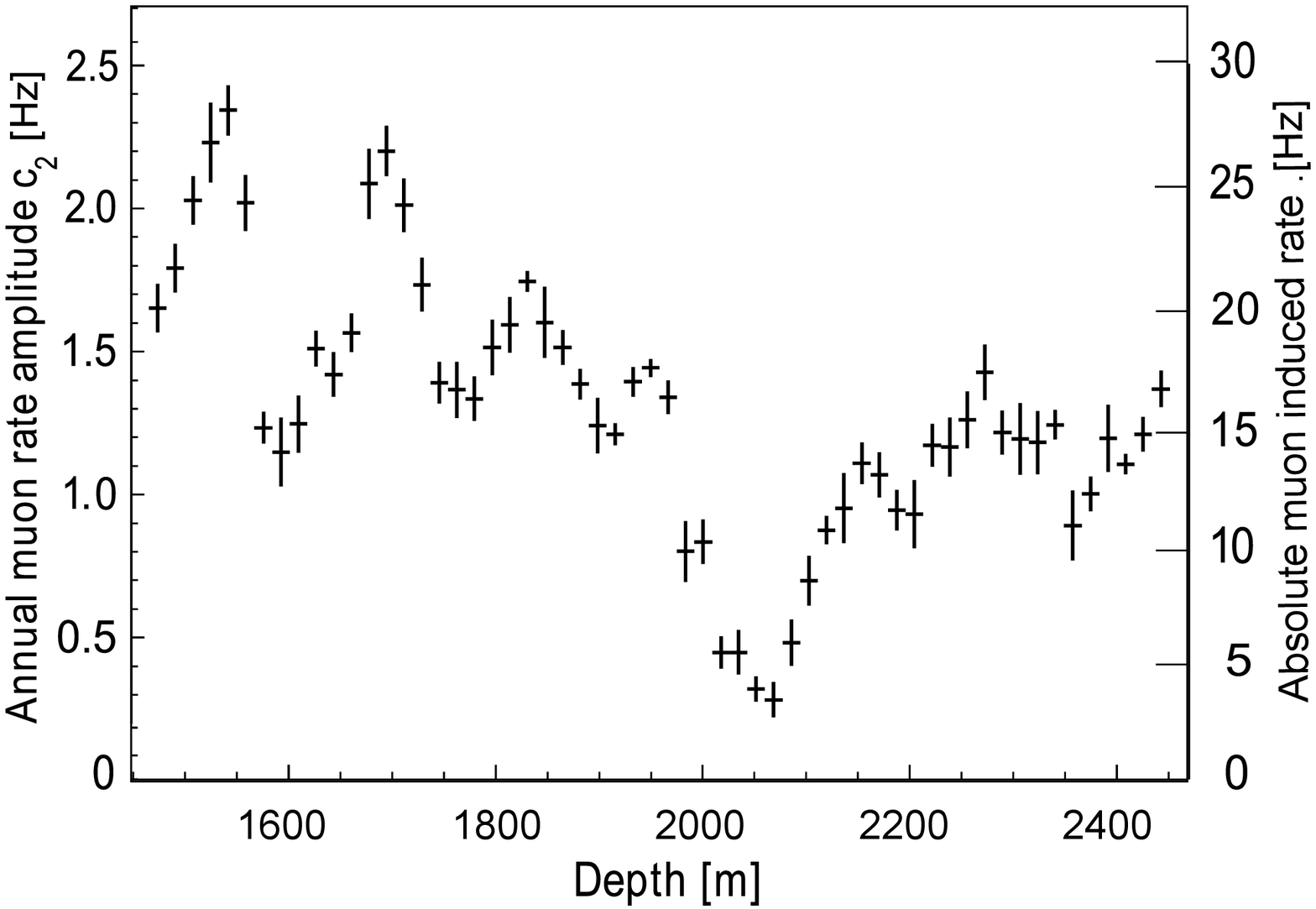}
\caption{Top: Rate of a typical DOM as function of time covering 556 days of lifetime as measured in \unit[0.5]{s} bins (baseline suppressed). The line corresponds to a rate fit according to Eq.~\ref{eq:exp+sin}. Bottom: Parameter $c_2$ and estimated muon induced rate as function of depth.  The variation with depth is mostly due to the optical properties of the ice and muons ranging out.}\label{fig:long_term_rates}
\end{figure}
\begin{figure}
\centering
\includegraphics[angle=0,width=0.45\textwidth]{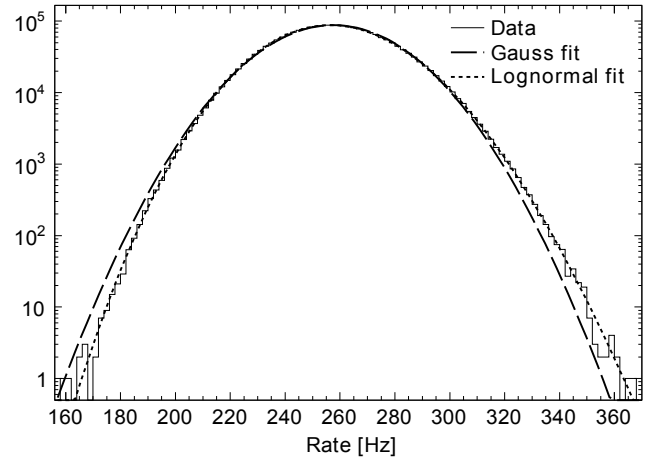}
\caption{Rate distribution of a typical standard efficiency DOM taken over 29 consecutive days. Each measurement corresponds to 0.5 s integration time. Gaussian
 and lognormal fits are shown. 
}\label{fig:rates}
\end{figure}

The DOM rates are characterized by an exponential rate decrease over long time periods and a slight seasonal modulation. For the purpose of this analysis, the formula 
\begin{equation}
	\label{eq:exp+sin}
	r(t) = r_0 + c_1 e^{-t/\tau} + c_2 \sin(2\pi (t/\mathrm{year}))
\end{equation}
represents the effects sufficiently well, as can be seen in Fig.~\ref{fig:long_term_rates}. The decay of the rate is likely due to a decrease of triboluminescence in the ice with time, a byproduct of the freezing process. For DOMs that have been in the ice for more than 3 years, the fall time $\tau$ exceeds 40 years, except for very deep DOMs where the freezing process takes longer ($\tau \approx 4$ years). In any case, the effect is negligible for the analysis which requires a stable rate within the analysis time window of \unit[10]{min}. The slightly skewed rate distribution of a single DOM is better described by a lognormal distribution than by a Gaussian (see Fig.~\ref{fig:rates} with \unit[250]{$\mu$s} dead time applied). The average rates for standard efficiency and high efficiency DOMs are determined to be $286\un{Hz}$ and $359 \un{Hz}$, respectively. Thanks to the tight quality control, variations between DOMs of the same type are small showing standard deviations of $26 \un{Hz}$ and $36\un{Hz}$, respectively. 
The seasonal DOM rate modulation is assumed to arise from a change in the atmospheric muon flux~\citep{tilav_icrc}. Fitting the time varying component, the parameter $c_2$ can be extracted (see Fig.~\ref{fig:long_term_rates}),
which tracks the effect of dust layers similarly to what was observed in the determination of effective volumes (see Fig.~\ref{veffplot}). If one interprets the effect as being due to stratospheric temperature variations measured to modulate the muon flux by $\Delta r\approx 8.3$\,\% in 2008, the averaged muonic contribution to single DOM rates is $c_2/\Delta r \approx 12 c_2 \approx \unit[16]{Hz}$. As will be discussed in the following section, statistical fluctuations in the atmospheric muon rate, despite the small average muon contribution to the DOM rate of  \unit[16]{Hz},
distort the significance spectrum considerably. 
\looseness-1
\subsection{Background Significance Distribution}
\label{sec:significance}
\begin{figure}[ht]
\centering
\includegraphics[angle=0,width=0.45\textwidth]{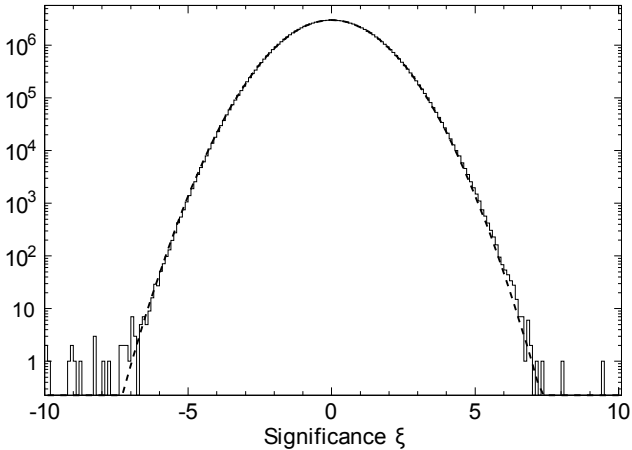}
\caption{Significance distribution in \unit[0.5]{s} binning for a detector uptime of 556 days with 22 and 40 strings deployed. The two outliers at $\xi$= 8 and 9.5 occured during test runs employing artificial light. The dashed line shows 
a Gaussian fit with $\sigma = 1.27$.
}\label{fig:significances}
\end{figure}

For purely uncorrelated background and high statistics, the significance $\xi$  is expected to be Gaussian distributed with width $\sigma = 1$ (see Sect.~\ref{sec:data_analysis}). 
As can be seen in Fig.~\ref{fig:significances}, the measured distribution is broader than expected and can be fairly well fitted by a Gaussian with width $\sigma = 1.27$. The broadening increases with the size of the detector and has reached $\sigma = 1.43$ with 79 operating strings. 
This broadening is due to non-Poissonian fluctuations in the number of hits deposited by atmospheric muons: highly energetic muons or muon bundles clustering in time leave correlated hits that will in general pass the $\chi^2$ cut. In the offline analysis, one can partly remove this effect as the number of muon induced coincident hits in neighboring DOMs is recorded for all triggered events. For the 79 (40) string configuruation, the broadening is thus reduced to $\sigma = 1.06$ (1.05). As this section describes the results of the online analysis, we do not apply this correction in the discussions below. 
 
An effective significance threshold of $\xi = 6.0$ provides an internal trigger for testing the system one to two times per day, while a threshold at $\xi = 7.1$ satisfies the SNEWS requirement of one false background trigger approximately once per 10 days. The Gaussian curve shown in Fig.~\ref{fig:significances} predicts one false background trigger within ten years at a threshold at $\xi = 11$. These thresholds are also depicted in Fig~\ref{fig:significance_vs_distance}. 
The entries at $\xi = 8$ and $\xi = 9.5$ are due to test runs with artificial light sources. 

\looseness-1
\subsection{Future improvements}
\label{sec:future_improve}Further optimizations may be applied to the data acquisition and analysis in the future, e.g. by incorporating a more sophisticated method to remove correlated noise, by excluding the bin-by-bin contribution of measured cosmic ray muon hits to the online rate measurement, 
by storing time stamps of all hits in case of a significant alarm to e.g. improve on the timing resolution and to track the average neutrino energy~\citep{bib:Baum, 
bib:Demiroers}, and by employing temporal templates in likelihood or cross-correlation studies. 
\subsection{Systematics}
\label{sec:systematics}
\begin{table*}[ht]
\caption{\label{tab:uncertainties}Summary of statistical and systematic uncertainties.}
\centering
\begin{tabular}{lccc}
\hline\hline

	Source of Systematics & \multicolumn{2}{c}{Uncertainties}\\
	                      & stat. (\% )& sys. (\%) \\												
	\hline
		Rate deviation in sliding average within 600 s buffer & 1.6 & - \\
		Ice density as function of depth & - & 0.2 \\
		Mean track length of electrons and positrons  $\bar x$ of $\mathrm{e^\pm}$ in ice & $\sim$\,0.9 & 5 \\
		Uncertainty of the effective volume $V^\mathrm{eff}_\gamma$ for photons & 1.3 & $\sim$\,12 \\
		Uncertainty on the effect of artifical dead time on the signal & - & 3 \\
		Uncertainty of the inverse beta-decay cross section  & - & $<$\,1 \\
		Uncertainty of the electron scattering cross sections & - & $<$\,1 \\
		Uncertainty of the oxygen scattering cross section & - & $\sim$\,[3,20] \\
		Angle dependent MSW oscillation in the mantle and core of the Earth & - &$\sim$\, [0,-8]\\
		\hline
\end{tabular}
\tablefoot{Estimated uncertainties on the total observed rate of all neutrino and anti-neutrino flavors. The earth effect strongly depends on the model and the neutrino angle of incidence. The uncertainty due to oxygen scattering depends on the neutrino energies and the neutrino channel. Assuming oxygen cross sections are known within a factor of two, rates may change between 3\,\% (Garching model) and 20\,\% (black hole model). For electron neutrino scattering alone, oxygen scattering accounts for about 40\,\% of the rate in the Garching model, with correspondingly large uncertainties.}
\end{table*}

There are three types of systematics relevant to this paper. The first type has to do with time dependent changes in the noise rate in all or a subset of DOMs that can mimic a supernova signal. These include high voltage variations, longterm trends such photomultiplier aging, weather effects and other experimental effecs. For the supernova monitoring, realtime analysis as well as triggering, longterm trends are accounted for by calculating the rate expectation values $\langle r_i\rangle$ and their standard deviation $\langle\sigma_i\rangle$ from a rolling average of the noise rate for each DOM over \unit[300]{s} on either side of the time of interest as was described in Sect.~\ref{sec:data_analysis}). The second type affects our understanding of the overall sensitivity of the detector. Ice properties, the wave length dependent quantum efficiency and the DOM thresholds all fall under this category. The third type is due to our current knowledge of relevant cross sections, the distance to supernovae, the neutrino-type dependent luminosity and energy, as well as oscillation effects in the star and in the Earth.

\subsubsection*{Detector Stability and Environmental Effects}
The detector behaves very stably under normal operation. Periods during drilling, tests with artificial light sources and periods with data acquisition problems as well as a few noisy modules are excluded (see Sect.~\ref{sec:qualification}) from the analysis. As discussed in~\ref{sec:long_term}, annual variations as well as shorter term modulations in the atmospheric muon flux change the observed rates which, however, are tracked by the rolling average. As hits from muons penetrating the detector are recorded simultaneously by the data acquisition system, they can be subtracted from the supernova rate measurements offline. Overall, the uncertainty on the supernova sensitivity associated with the detector stability is estimated to be small (1.6\,\%). 
 
The data were checked for other external sources of rate changes such as seismic activity and varying magnetic or electric fields as tracked by magnetometers and riometers at the South Pole.
Only magnetic field variations show a slight, albeit insignificant, influence on the rate deviation of \unit[$-1.3 \cdot 10^{-6}$]{Hz/nT}. The influence is 30 times lower in IceCube than in AMANDA due to a wire mesh $\mu$ metal shielding in IceCube DOMs. 

\subsubsection*{Ice Properties and Sensitivity of the Detector}
Dust and air bubbles in the natural ice medium cause photons to scatter, while dust and the ice itself determine the absorption length. The range of ice densities  $\rho_{ice}=(919.6\, \pm\,  1.6)$ kg/m$^3$~\citep{PRICE} reflects the 0.4\,\%  density decrease due to the temperature increase between (1.4--2.4) km depth. 
Scattering dominates in the shallow ice above \unit[1400]{m} and possibly in the ice of the hole around the DOMs, which refreezes soon after deployment.  
Uncertainties in the optical properties of the hole ice are estimated to affect the effective volume determination by $<$ 1\,\%. More important are the uncertainties in the description of ice properties of the Antarctic glacier.

The distributions of the photon arrival times and number of photons received at AMANDA modules from artificial light sources were used to derive the scattering length at different depths. Pulsed and continuous LED and laser sources give complementary measurements. The measurements of ice properties are consistent to within 6\,\% of each other including statistical and systematic uncertainties both for the scattering and absorption measurements. The information on photon propagations is stored in tables contributing an estimated 1\,\% uncertainty due to the finite binning.  Our knowledge of ice properties and corresponding simulation methods continue to improve. Variation of DOM optical sensitivites and effects of the photomultiplier threshold on single photo-electron pulses lead to an $\approx$\,10\,\% uncertainty.
We assumed a $(7\pm 3)$\,\% loss of light due to cables that shadow the photomultiplier surface. The effects discussed in this paragraph add up to an overall 12\,\% uncertainty. 

The track length of a positron or electron, including that of secondaries with kinetic energies above the Cherenkov threshold of 0.272 MeV, depends linearly on the initial lepton energy. The statistical uncertainty of the GEANT-4 calculation, including a systematic difference between electrons and positrons, is 0.3\,\%. The implementations of low energy electromagnetic processes have been cross checked between GEANT and NIST ESTAR-ICRU37 compilations. Good agreement has been found in particular for electron ranges~\citep{Amako}. NIST quotes a (2 -- 5)\,\% systematic uncertainty on their implementation of electromagnetic cross sections in the energy range relevant to supernovae.  
Event to event statistical fluctuations in the track length and in the number of Cherenkov photons ($\approx$ 2\,\%) are negligible when investigating the ensemble of all DOMs.  

\subsubsection*{External Sources of Systematics}
\label{sec:theory Systematics}

The estimated uncertainties of the cross sections are listed in Table~\ref{tab:cross}. Those associated with oxygen scattering processes are large and difficult to assess. Due to the strong energy dependence of $^{16}$O neutrino cross sections, the impact of this uncertainty depends on the energy spectra of particular models and the assumed oscillation scenarios. Processes involving only $\nu_\mathrm{e}$ scattering are particularly affected.	
The total systematic uncertainty from detector effects and cross sections on the total rate of all neutrinos (electron neutrinos) is $\approx$ 14\,\% (25\,\%). 

The distance to stars in our Galaxy is typically known to 25\,\% accuracy~\citep{Scheffler}. However, the distance of a supernova can in principle be measured by interpreting its light curve with an accuracy of (5 -- 10)\,\%~\citep{eastman}. Unfortunately, a considerable fraction of supernovae occurring in our Galaxy may be obscured by dust at optical wavelengths.

The uncertainties in the supernova collapse models are large and difficult to assess. The $\nu_\mathrm{e}$ rate from the neutronization burst is largely independent of the progenitor mass;  the corresponding uncertainties are estimated to be around 10\,\%; uncertainties arising from neutrino oscillations are estimated to be below 5\,\% for a normal hierarchy~\citep{PhysRevD.71.063003}.

Oscillations in the Earth strongly depend on the incoming neutrino direction and may lead - depending on the neutrino hierarchy - to a maximal rate decrease of  8\,\% and 3\,\% during the cooling phases of the Lawrence Livermore and Garching models, respectively. The differences between various oscillation scenarios may be as large as 30\,\% or even 50\,\% in the case of black hole formation. 

\section{Performance Simulations}\label{sec:physicsperf}
\label{sec:I3performance}
In this section we will discuss the capability of IceCube to characterize details of the core collapse of massive stars and of the supernova remnant, as well as the insights IceCube may provide into the properties of neutrinos and their interactions. There remain significant uncertainties in our understanding of the neutrino emission from supernova explosions, necessitating comparisons between several models to map the parameter space. In order to illustrate IceCube's performance, we will refer to specific models chosen to span the possible range of supernova progenitor masses and neutrino energy spectra. We will also refer to more speculative models in order to demonstrate IceCube's high statistical precision in the detection of modulations of the neutrino light curve from astrophysical effects. 

When discussing the complete accretion and cooling phase extending to \unit[15]{s}, we refer to recent O-Ne-Mg core models~\citep{Huedepohl} and the older Lawrence-Livermore model~\citep{AstropPhys.496.216} as examples with low and high progenitor star masses. The calculations consider only the radial dimension as a parameter. 
When discussing the first 800 milliseconds of the burst we also refer to the Garching model~\citep{AstrAstrop.450.1} as an example for calculations with sophisticated transport mechanisms.
Because it assumes only half of the initial star mass, \unit[(8 -- 10)]M$_\odot$ instead of \unit[20]{M$_\odot$}, it predicts fewer neutrinos than the Lawrence-Livermore model.

In order to be compatible with other studies, we will usually show experimentally predicted neutrino light curves for distances of 10 kpc, roughly corresponding to the center of our Galaxy. Depending on the model for the supernova precursor distribution, between 44\,\%~\citep{Ahlers} and 53\,\% \citep{Bahcall:1982fx} of all core collapse candidate stars in the Milky Way are expected to occur within this distance.  About 90\,\% of all supernovae are predicted to occur within 15.4 kpc~\citep{Mirizzi} to 17.5 kpc~\citep{Ahlers} distance from the Earth. 

In the study of star matter oscillation effects, we restrict ourselves to the comparisons of the three scenarios A-C for neutrino hierarchy and $\theta_{13}$ mixing angles that were introduced in Sect.~\ref{sec:pheno}. For some comparisons, we also show distributions with star matter oscillations turned off. 

All simulations are performed for the final IceCube array with 4800 standard and 360 high efficiency DOMs. We assume that 2\,\% of the DOMs are excluded from the analysis, either because they are not working or they give unstable rates.  
The background noise was accounted for in two different way. For the determination of the significance and galaxy coverage, the simulated signal was randomized assuming a Poissonian distribution and added to noise data taken from experimental measurements and analyzed with the real-time reconstruction programs. For the simulation and comparision of various models, we added the calculated and randomized signal rates to the noise of the floor drawn from a Gaussian with mean value and standard deviation derived from data. 

Due to correlated pulses from radioactive decays and atmospheric muons, the measured sample standard deviation in data taken with 79 strings is $\approx 1.3$ and $\approx 1.7$ times larger than the Poissonian expectation for 2 ms and 500 ms bins, respectively. It is possible to subtract roughly half of the hits introduced by atmospheric muons from the total noise rate in the offline analysis, as the number of coincident hits in neighboring DOMs is recorded for all triggered events. We apply this correction to all Monte Carlo analyses described in this section, as this procedure lowers the standard deviation to 
$(1.24 -1.32)\sqrt{\sum_i r_i}$, slightly dependent on the binning.   

Unless noted otherwise, we will use a likelihood ratio method to determine the range within which models can be distinguished. From sets of several thousand test experiments, we will typically determine limits at the 90\,\% confidence level, while requiring that the tested scenario is detected in at least 50\,\% of the cases. Note that the ranges obtained should be interpreted as optimal as we assume that the model shapes are perfectly known and only the overall flux is left to vary; we also disregard the possibility that multiple effects, such as matter induced neutrino oscillations and neutrino self-interactions, could co-exist and thus may be hard to disentangle.  

\subsection{Expected Supernova Signal}
\label{sec:I3expectedsignal}
\begin{figure}[htb]
\centering
\includegraphics[angle=0,width=0.45\textwidth]{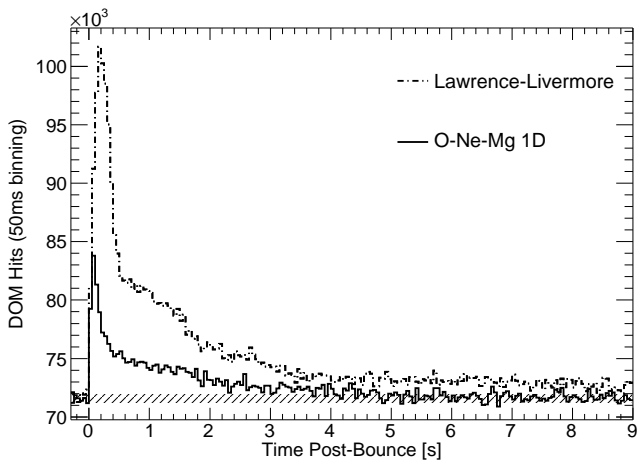}
\caption{Expected rate distribution at \unit[10]{kpc} distance for the Lawrence-Livermore model (dashed line) and O-Ne-Mg model by~\cite{Huedepohl} with the full set of neutrino opacities (solid line). . 
The $1\,\sigma$-band corresponding to measured detector noise (hatched area) has a width of about $\pm$\,\unit[330]{counts}.} 
\label{fig:rate0}
\end{figure}

Evaluating Eq.~\ref{eq:rate} one obtains the rate spectra of Fig.~\ref{fig:rate0} for a supernova at \unit[10]{kpc} distance. 
With a maximal signal-over-noise ratio of $\approx 55$ for the Lawrence-Livermore model, the neutrino burst can clearly be detected with IceCube. Also, the still hypothetical accretion phase lasting from \unit[(0 - 0.5)]{s} can be separated from the subsequent cooling phase with high statistical precision. The study of the cooling phase is limited by the photomultiplier noise in particular for the case of the light O-Ne-Mg model by~\cite{Huedepohl}.

The oscillation scenario B for an inverted neutrino mass hierarchy shows the largest signal for the Lawrence-Livermore and Garching models because energetic $\bar\nu_\mathrm{x}$ will oscillate into $\bar\nu_\mathrm{e}$, harden their spectrum and thus increase the detection probability.  
The scenario without any oscillation is presented as a reference and leads to the weakest signal. Scenario A (normal hierarchy) and Scenario C (very small $\theta_{13} < 0.09^\circ$)
 are hard to distinguish due to their very similar effect on neutrino mixing. 
\begin{figure*}[ht]
\centering
\includegraphics[angle=0,width=0.48\textwidth,bb=50 -50 600 380]{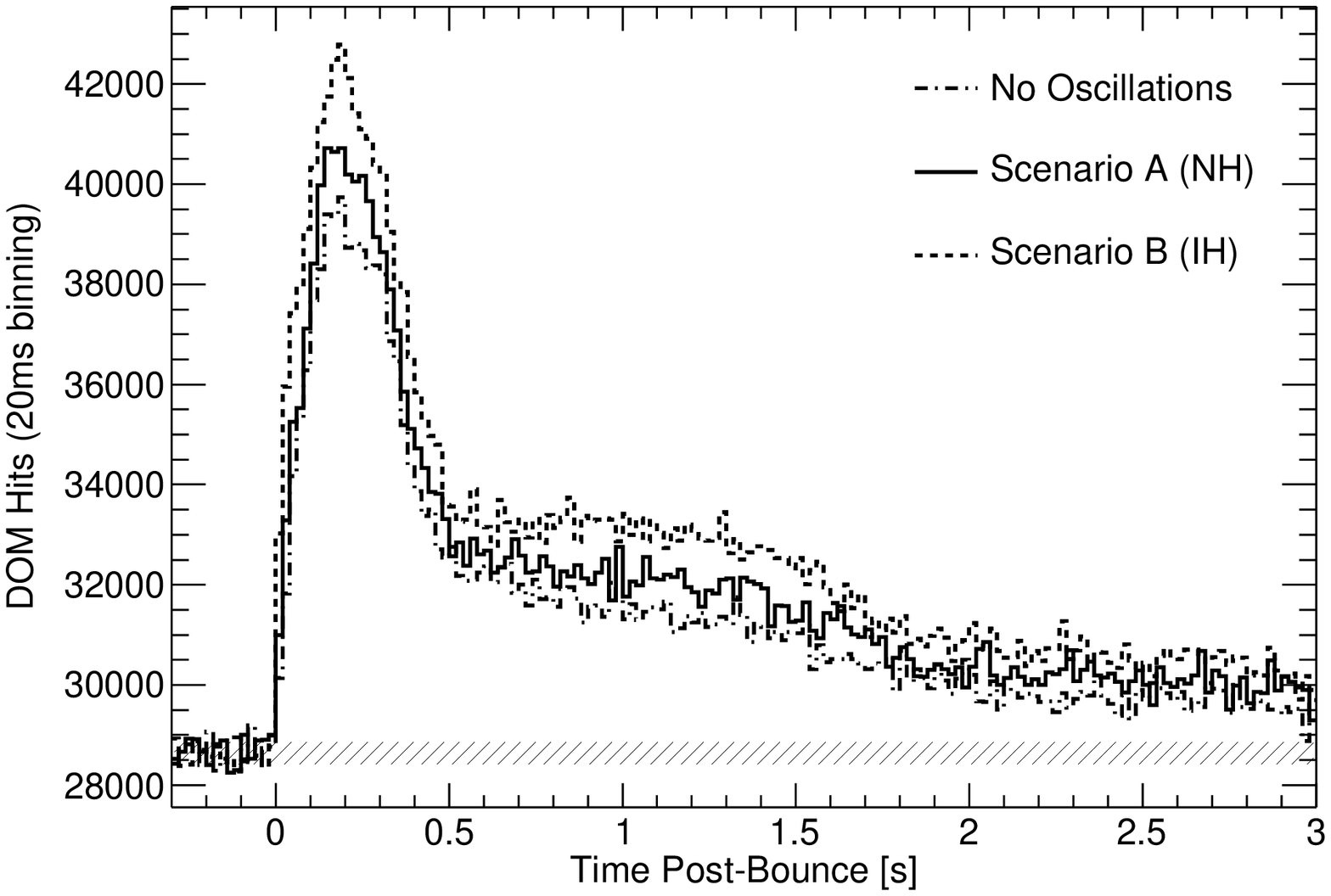}
\includegraphics[angle=0,width=0.48\textwidth,bb=-20 -50 530 380]{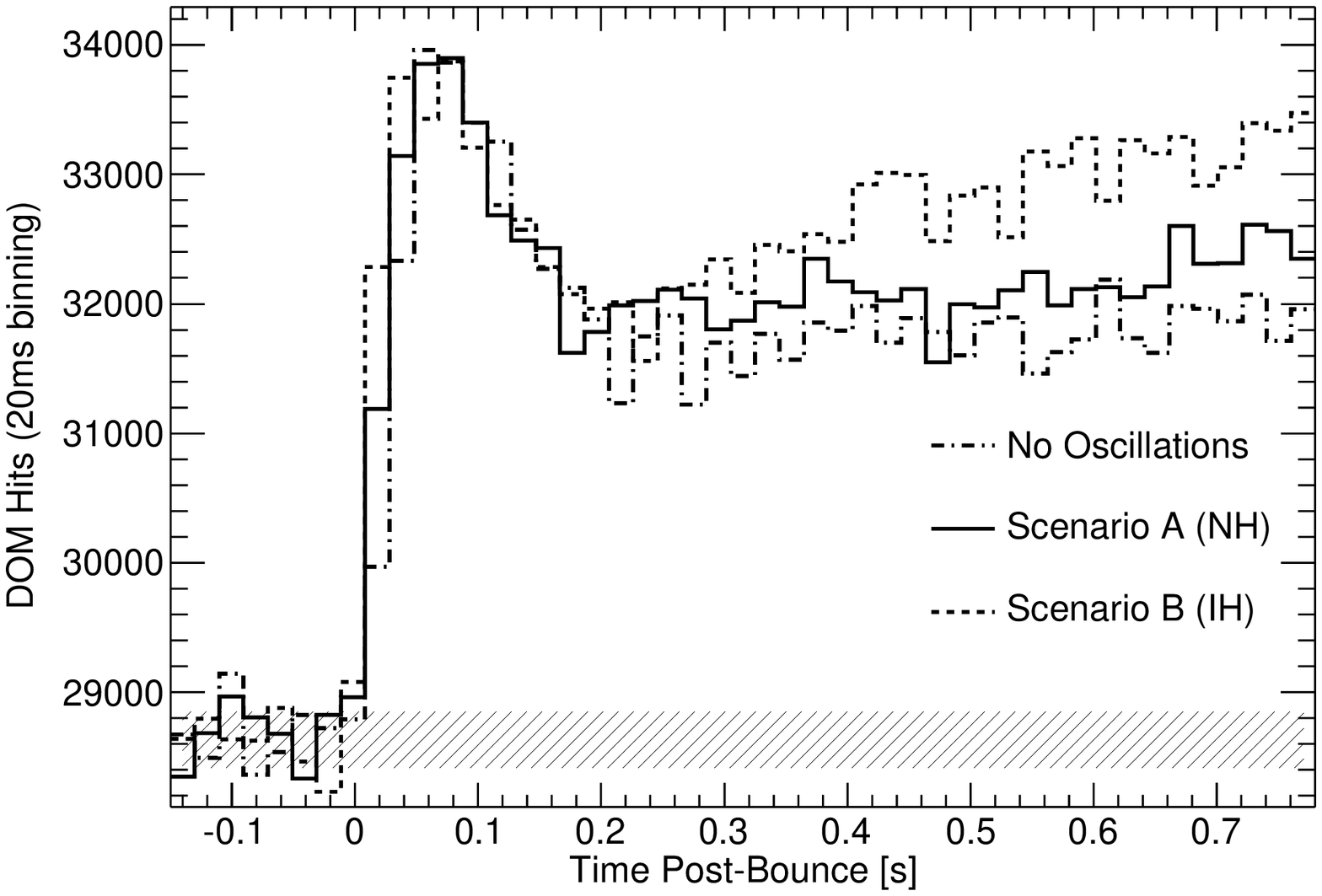}
\includegraphics[angle=0,width=0.48\textwidth,bb=10 15 560 395]{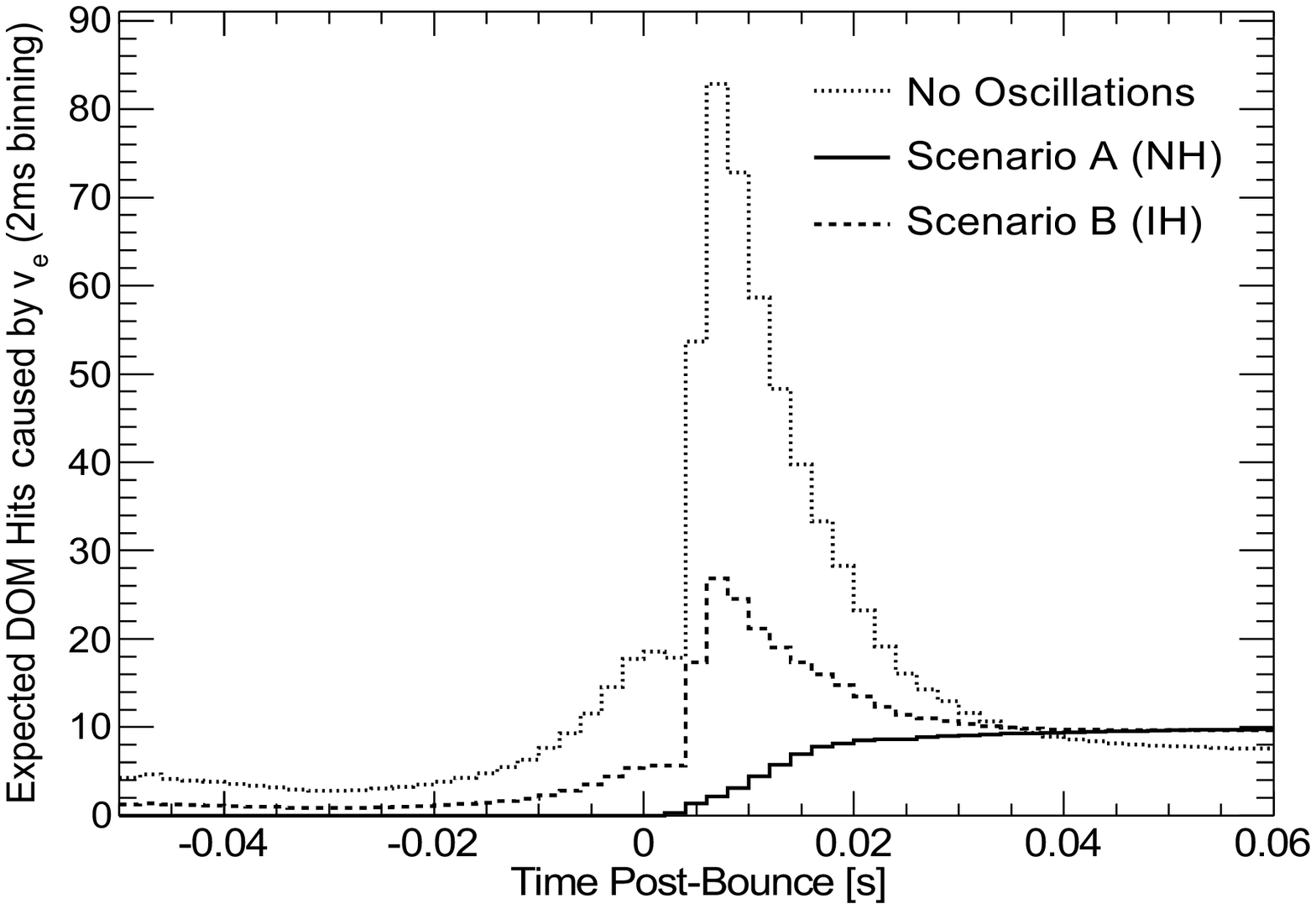}
\includegraphics[angle=0,width=0.48\textwidth,bb=0 0 550 380]{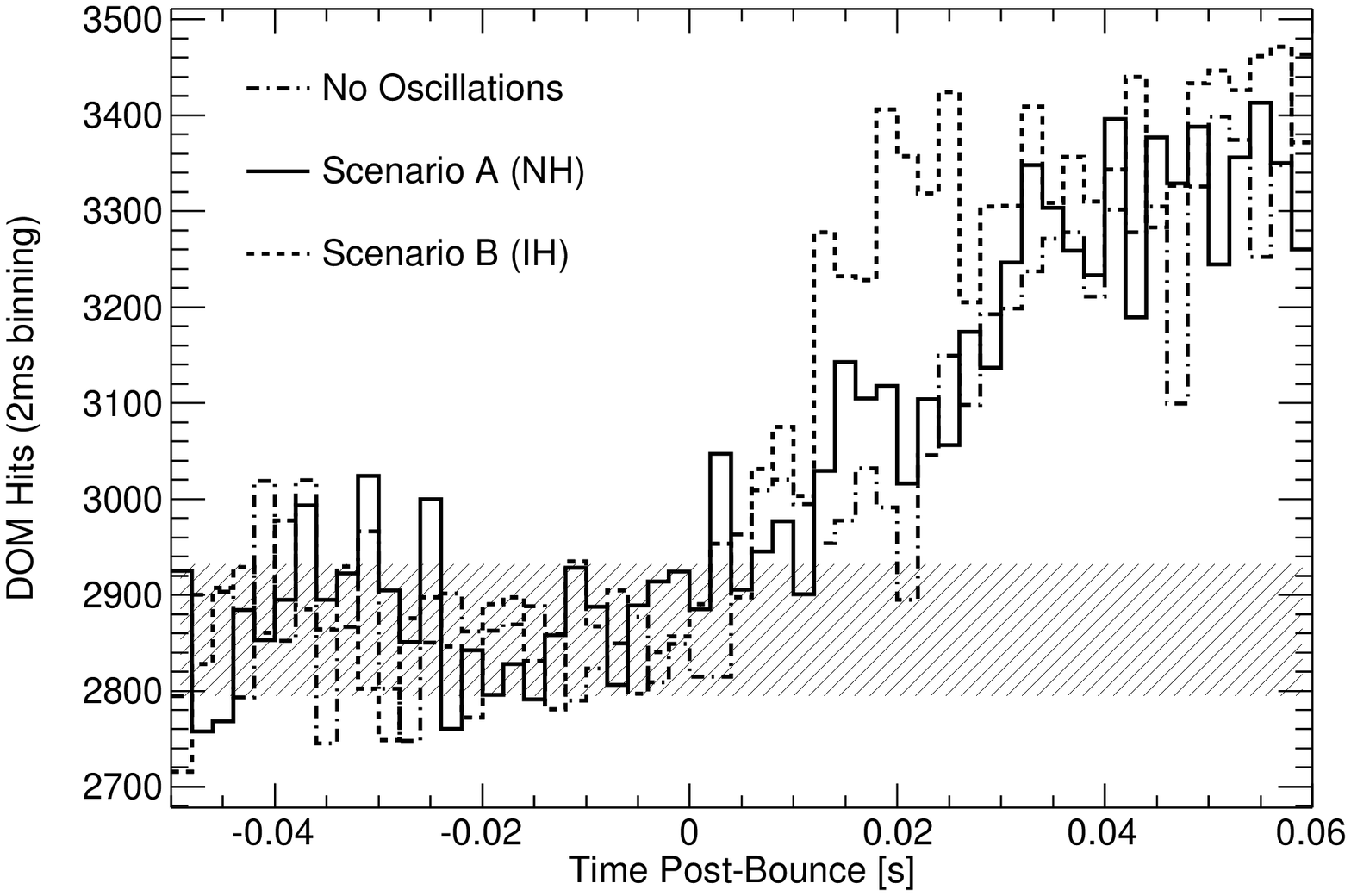}
\caption{Top: Expected rate distribution at \unit[10]{kpc} supernova distance for oscillation scenarios A (normal hierarchy) and B (inverted hierarchy). Fluxes and energies in the left plot are taken from the Lawrence-Livermore model and in the right plot from the Garching model using the equation of state of~\citet{Lattimer}. 
 Scenario C (not shown) is almost indistinguishable from Scenario A. The case of no oscillation is given as a reference. 
Bottom: Expected average signal rate distribution at \unit[10]{kpc} distance in finest \unit[2]{ms} binning for Scenarios A and B  using the Garching model; the unlikely case of no oscillation is given as a reference. The left plot shows the expected $\nu_\mathrm{e}$ induced signal. As can be seen from the right plot, the signal is no longer apparent, once the large contribution due to the inverse beta decay and the expected DOM noise are added. 
The $1\,\sigma$-bands corresponding to measured detector noise (hatched area) have a width of about  $\pm$\, \unit[215]{counts} for a \unit[20]{ms} binning and $\pm$\,\unit[70]{counts} for a \unit[2]{ms} binning.}
\label{fig:rate}
\end{figure*}

Clear differences between the oscillation scenarios in absolute rate and shape appear in Fig.~\ref{fig:rate}. Assuming that the model shapes are known but not necessarily the overall normalization, the inverted hierarchy can be distinguished from  the null hypothesis of a normal hierarchy up to distances of 16 kpc. 

\subsection{Significance and Galaxy Coverage}
\label{sec:I3significance}
\begin{figure}[htb]
\centering
\includegraphics[angle=0,width=0.45\textwidth]{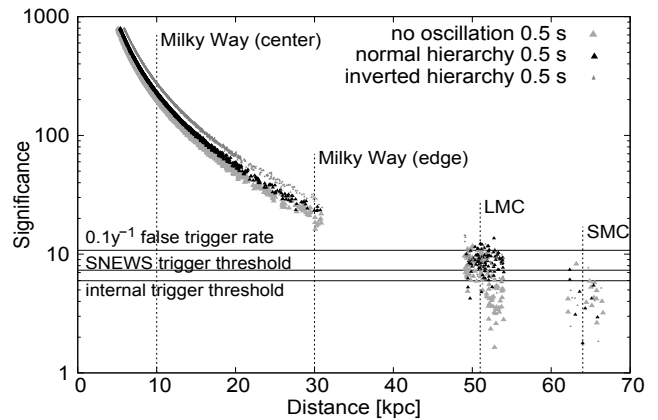}
\caption{Significance versus distance 
assuming the Lawrence-Livermore model. The significances are increased by neutrino oscillations in the star by typically 15\,\% in case of a normal hierarchy (Scenario A) and 40\,\% in case of an inverted hierarchy (Scenario B).  
The Magellanic Clouds as well as center and edge of the Milky Way are marked. The density of the data points reflect the star distribution.}\label{fig:significance_vs_distance}
\end{figure}

The simulation of an expected signal from a supernova within the Milky Way has to take into account the number of likely progenitor stars in the Galaxy as a function of the distance from Earth. The expected significances of supernova signals according to the Lawrence-Livermore model for three oscillation scenarios are shown in Fig.~\ref{fig:significance_vs_distance}. For this particular model, the significances for the \unit[4]{s} and \unit[10]{s} binning turn out to be approximately 20\,\% and 50\,\% lower than for
\unit[0.5]{s}, respectively. 
For the graph, the supernova progenitor distribution predicted by ~\citet{Bahcall:1982fx} was used. For the Magellanic Clouds, which contain roughly 5\,\% of the stars in the Milky Way, a uniform star distribution along the diameters of the galaxies was assumed for simplicity. 

IceCube is able to detect supernovae residing in the Large Magellanic Cloud (LMC) with an average significance of $(5.7 \pm 1.5 )$ $\sigma$ in a \unit[0.5]{s} binning, assuming the Lawrence-Livermore model. The uncertainty reflects different oscillation scenarios. Supernovae in the Small Magellanic Cloud (SMC) can be detected with an average significance of $(3.2 \pm 1.1)$ $\sigma$ and will in general not trigger sending an alarm to SNEWS, as indicated by a horizontal line in Fig.~\ref{fig:significance_vs_distance}. IceCube will observe supernovae in the entire Milky Way with at least a significance of 12 $\sigma$ at \unit[30]{kpc} distance. 

\subsection{Onset of Neutrino Production}

The analysis of the deleptonization peak that immediately follows the collapse is of considerable interest, since its magnitude and time profile are rather independent of the initial star mass and of the nuclear equation of state; the variation is estimated by~\citep{keil} to be around 6\,\%. Thus the electron neutrino luminosity 
may be used as a standard candle to measure the distance to the supernova.

As the deleptonization peak lasts for only \unit[10]{ms}, the data are evaluated in the finest available time binning of \unit[2]{ms}, as depicted in Fig.~\ref{fig:rate}. The deleptonization signal is detected by the elastic $\nu_\mathrm{e} + \mathrm{e^-}\rightarrow \nu_\mathrm{e} + \mathrm{e^-}$ reaction with a cross section times the number of targets $\approx 50$ times smaller than for the $\bar{\nu}_\mathrm{e} + \mathrm{p} \rightarrow \mathrm{e^+} + \mathrm{n}$ interaction. As the $\bar{\nu}_\mathrm{e}$ flux rises rapidly following the collapse, the deleptonization peak remains almost completely hidden, especially when neutrinos oscillate in the star. In this case the subtle structure may be resolved only for distances $d \leq \unit[2]{kpc}$. 

Largely independently of the model, each oscillation scenario shows a characteristic slope of the rate increase around the deleptonization peak. Quantifying this by a series of several thousand simulations for the Garching and Lawrence-Livermore models and considering oscillation Scenarios A-C and the case of no oscillation, it is possible to establish the inverse hierarchy (Scenario B) w.r.t. the normal hierarchy (Scenario A)  with 90\,\% C.L. for distances $d \leq \unit[6]{kpc}$ (corresponding to 21\,\% of all progenitor stars, when accounting for the effect of spiral arms~\citep{Ahlers}). 

\subsection{Shock Waves}

\begin{figure}[ht]
\centering
\includegraphics[angle=0,width=0.45\textwidth]{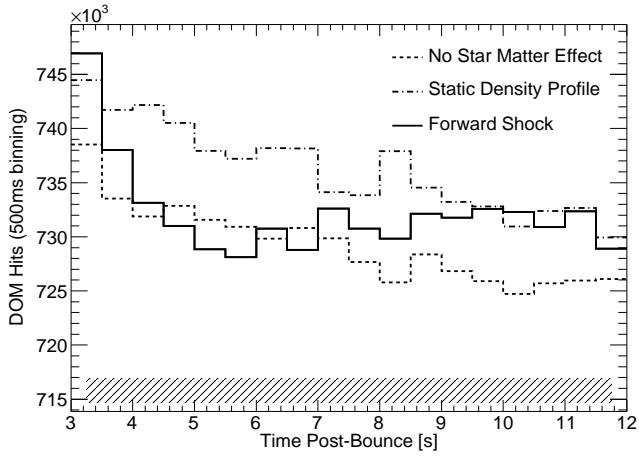}
\caption{Effect of a forward moving shock wave applied to supernova at 10 kpc distance, modelled according to the Lawrence-Livermore model assuming an inverted hierarchy and $\theta_{13} > 0.9^\circ$
was assumed. A forward shock wave can be distinguished from a static density profile and the case of no star matter effect. The $1\,\sigma$-band corresponding to measured detector noise (hatched area)  has a width of about $\pm$\,\unit[1150]{counts}.  
}\label{fig:instabilities}
\end{figure}

For an inverted hierarchy (Scenario B), the rate distribution should reveal the effects of forward and backward moving shock waves traveling through the collapsing star during the cooling phase, (3 -- 10) s after bounce. Assuming the specific model of~\citet{Tomas} (see Fig.~\ref{fig:instabilities}), scenarios with a static density profile and one forward shock wave can be distinguished at 90\,\% C.L. up to distances of 13 kpc; the distance reduces to 10 kpc in a scenario with one forward and one reverse shock wave (not shown).  

\subsection{Quark Star and Black Hole Formation}
\begin{figure}[ht]
\centering
\includegraphics[angle=0,width=0.48\textwidth]{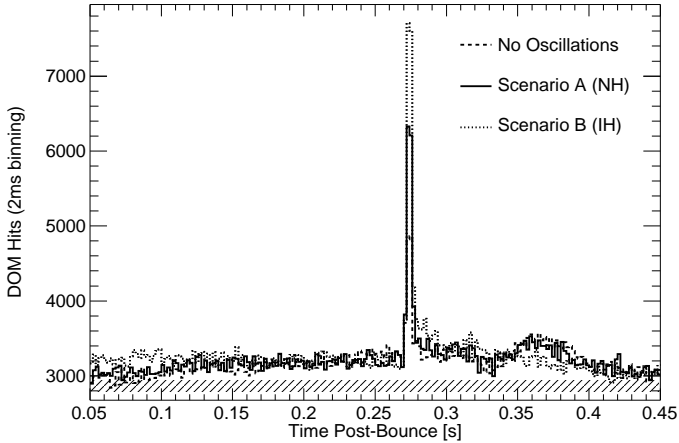}
\caption{Comparison of the neutrino light curve with quark-hadron phase transition for a 10 $M_\odot$ progenitor at 10 kpc distance. Three neutrino oscillation scenarios are shown (see legend). The observation of the sharp $\bar\nu_\mathrm{e}$ induced burst, 257 ms $< t< \, $261 ms after the onset of neutrino emission, would constitute direct evidence of quark matter. The hatched $1\,\sigma$-band corresponding to detector noise has a width of about $\pm$\,\unit[70]{counts}. }\label{fig:quark}
\end{figure}

\begin{figure}[ht]
\centering
\includegraphics[angle=0,width=0.46\textwidth, bb= 24 7 555 379]{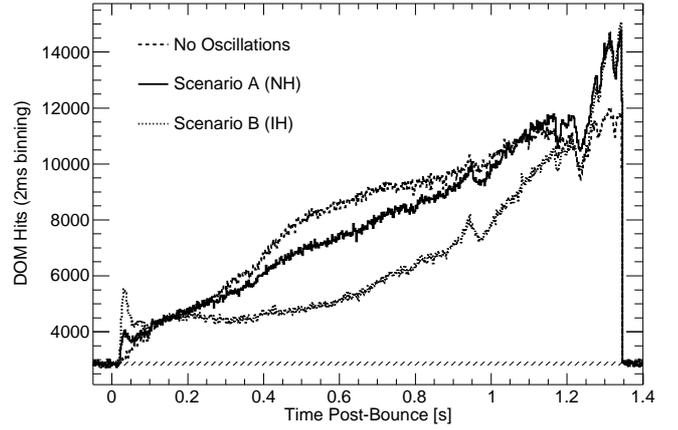}
\caption{Expected neutrino signal from the gravitational collapse of a non rotating massive star of 40 solar masses into a black hole at 10 kpc distance for a hard equation of state~\citep{Shen} following~\citet{Sumiyoshi}. The $1\,\sigma$-band corresponding to detector noise (hatched area) has a width of about $\pm$\,\unit[70]{counts}. }\label{fig:bh}
\end{figure}

IceCube is particularly well suited to study fine details of the neutrino flux as function of time. As an example, Fig.~\ref{fig:quark} shows a simulation based on the prediction of~\citet{Dasgupta-2010} for the formation of a quark star. The model predicts a sudden spike in the $\bar\nu_\mathrm{e}$ flux lasting for a few ms while the neutron star turns to a quark star; the time of the QCD phase transition can be determined with sub-ms accuracy. The likelihood ratio test gives a deviation larger than 5~$\sigma$ from the hypothesis of no quark star formation for distances up to 30 kpc.  Height and shape of the peak depend on the neutrino hierarchy. Scenarios A and B can be distinguished at 90\,\% C.L. up to distances of 30 kpc. 
\vfill
Fig.~\ref{fig:bh} shows a simulation based on the prediction of~\citet{Sumiyoshi} for the formation of a black hole following a collapse of a 40 solar mass progenitor star. Neutrinos  reach energies up to 27 MeV ($\nu_\mathrm{e}$ and $\bar{\nu}_\mathrm{e}$) and 40 MeV ($\nu_{\mu}$ and $\nu_{\tau}$), carry a correspondingly large detection probability and thus produce very clear evidence for the formation of the black hole after 1.3 s, when the neutrino emission is expected to fade exponentially (not realized in the simulation). For Fig.~\ref{fig:bh}, a hard equation of state~\citep{Shen} was chosen, leading to black hole formation after 1.3 s. This corresponding drop can be identified at higher than 90\,\% C.L. for all stars in our Galaxy and the Magellanic Clouds. 
\subsection{Neutrino Hierarchy Sensitivity and Rate Summary}
\begin{figure}[ht]
\centering
\includegraphics[angle=0,width=0.45\textwidth,bb=  0 0 564 531]{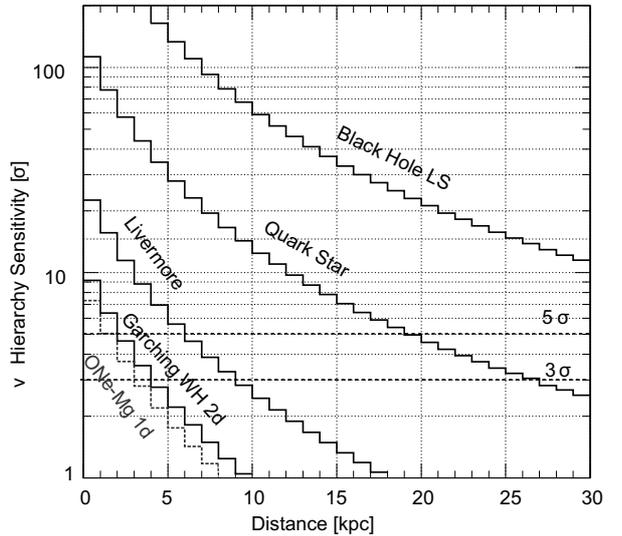}
\caption{Number of standard deviation with which scenarios A (normal hierarchy) and B (inverted hierarchy) can be distinguished in at least 50\,\% of all cases as function of supernova distance for some of the models listed in Table~\ref{tab:eventsummary}. A likelihood ratio method was used assuming known model shapes.}\label{fig:reach}
\end{figure}

\looseness -1
The number of standard deviation with which normal and inverted $\nu$ hierarchies (Scenarios A and B) can be distinguished are plotted in Fig.~\ref{fig:reach} as function of the supernova distance for selected models. The values represent the optimal cases when model shapes (but not necessarily the absolute fluxes) are perfectly known. 
Table~\ref{tab:eventsummary} lists the number of neutrino induced photon hits that would be recorded by IceCube on top of the DOM noise for various supernova models. Note that the number of expected signal hits scales with 1/distance$^2$; the dependence of the detection significance as function of distance can be read from Fig.~\ref{fig:significance_vs_distance}.
\begin{table*}[ht]
\caption{\label{tab:eventsummary}Expected rates.}
\centering
\begin{tabular}{llccc}
\hline\hline
  Model & Reference & Progenitor       & $\#\nu$'s  & $\#\nu$'s   \\
      &           & mass ($M_\odot$) & $t<380$ ms & all times \\ 
\hline
``Livermore''                   & \citep{AstropPhys.496.216}						& 20        &$0.174\times 10^6$& $0.79\times 10^6$\\
``Garching LS-EOS 1d''          & \citep{AstrAstrop.450.1}   & $8-10$    &$0.069\times 10^6$ & -               \\
``Garching WH-EOS 1d''          & \citep{AstrAstrop.450.1}   & $8-10$    &$0.078\times 10^6$ & -               \\
``Garching SASI 2d''            & \citep{Marek}              & 15        &$0.106\times 10^6$ & -               \\
``1987A at 10 kpc''             & \citep{Pagliaroli-b}   & $15-20$   & &$(0.57\pm 0.18)\times 10^6$        \\
``O-Ne-Mg 1d''                  & \citep{Huedepohl}          & 8.8       &$0.054\times 10^6$ &$0.17\times 10^6$\\
``Quark Star (full opacities)'' &\citep{Dasgupta-2010}       &  10       &$0.067\times 10^6$ & -               \\
``Black Hole LS-EOS''           &\citep{Sumiyoshi}           &  40       &$0.395\times 10^6$ &$1.03\times 10^6$ \\
``Black Hole SH-EOS''           &\citep{Sumiyoshi}           &  40       &$0.335\times 10^6$ &$3.40\times 10^6$ \\
\hline
\end{tabular}
\tablefoot{Number of recorded DOM hits in IceCube ($\approx \# \nu$'s) for various models of the supernova collapse and progenitor masses assuming a distance of 10 kpc, approximately corresponding to the center of our Galaxy. A normal neutrino hierarchy is assumed.}
\end{table*}

\section{Conclusion}\label{sec:conclusions}
A high statistics observation of the supernova neutrino flux would provide valuable information on astrophysics and the properties of neutrinos.
 IceCube was completed in December 2010 and monitors $\approx$ 1 km$^3$ of deep Antarctic ice for particle induced photons with 5160 photomultiplier tubes. 
Since 2009 it supersedes AMANDA in the SNEWS network. With a 250 $\mu$s artificial dead time setting, the average DOM noise rate is 286 Hz. The rates remain constant over time with a small modulation induced by changes in the atmospheric muon flux; they hardly vary across the detector once the DOMs have been frozen in for a sufficiently long period. The data taking is very reliable and covers the whole calendar year, including periods when new strings were deployed. The uptime has continuously improved toward a goal of $>98$\,\% and reached 96.7\,\% in 2009. IceCube's sensitivity corresponds to a megaton scale detector for galactic supernovae, triggering on supernovae with about 200, 20, and 6 standard deviations at the galactic center (10 kpc), the galactic edge (30 kpc), and the Large Magellanic Cloud (50 kpc).  IceCube cannot determine the type, energy, and direction of individual neutrinos and the signal is extracted statistically from rates that include a noise pedestal. On the other hand, IceCube is currently the world's best detector for establishing subtle features in the temporal development of the neutrino flux. The statistical uncertainties at $10\un{kpc}$ distance in $20\un{ms}$ bins around the signal maximum are about 1.5\,\% and 3\,\% for the Lawrence Livermore and Garching models, respectively. 

Depending on the model, in particular the progenitor star mass, the assumed neutrino hierarchy and neutrino mixing, the total number of recorded neutrino induced photons from a burst 10 kpc away ranges between $\approx\, 0.17\times 10^6$  (8.8 M$_{\odot}$ O-Ne-Mg core), $\approx 0.8\, \times 10^6$ (20 M$_{\odot}$ iron core) to $\approx 3.4\, \times 10^6$ for a 40 M$_{\odot}$ progenitor turning into a black hole. 
For a supernova in the center of our Galaxy, IceCube's large statistics would allow for a clear distinction between the accretion and cooling phases, an estimation of the progenitor mass from the shape of the neutrino light curve, and for the observation of short term modulation due to turbulent phenomena or forward and reverse shocks during the cooling phase. The deleptonization peak associated with the neutron star formation, however, may be hard to observe since the electron neutrino cross section in ice is small. 
IceCube will be able to distinguish inverted and normal hierarchies for the Garching, Lawrence-Livermore and black hole models for a large fraction of supernova bursts in our Galaxy provided that the model shapes are known and  $\theta_{13} > 0.9^\circ$. The slope of the rising neutrino flux following the collapse can be used to distinguish both hierarchies in a less model dependent way for distances up to 6 kpc at 90\,\% C.L. 
As in the case of the inverted hierarchy, coherent neutrino oscillation will enhance the detectable flux considerably. A strikingly sharp spike in the $\bar{\nu}_\mathrm{e}$ flux, detectable by IceCube for all stars within the Milky Way,  would provide a clear proof of the transition for neutron to a quark star as would be the sudden drop of the neutrino flux in case of a black hole formation.  

\begin{acknowledgements}
We acknowledge the support from the following agencies:
U.S. National Science Foundation-Office of Polar Programs,
U.S. National Science Foundation-Physics Division,
University of Wisconsin Alumni Research Foundation,
the Grid Laboratory Of Wisconsin (GLOW) grid infrastructure at the University of Wisconsin - Madison, the Open Science Grid (OSG) grid infrastructure;
U.S. Department of Energy, and National Energy Research Scientific Computing Center,
the Louisiana Optical Network Initiative (LONI) grid computing resources;
National Science and Engineering Research Council of Canada;
Swedish Research Council,
Swedish Polar Research Secretariat,
Swedish National Infrastructure for Computing (SNIC),
and Knut and Alice Wallenberg Foundation, Sweden;
German Ministry for Education and Research (BMBF),
Deutsche Forschungsgemeinschaft (DFG),
Research Department of Plasmas with Complex Interactions (Bochum), Germany,
Research Center Elementary Forces and Mathematical Foundations (Mainz), Germany;
Fund for Scientific Research (FNRS-FWO),
FWO Odysseus programme,
Flanders Institute to encourage scientific and technological research in industry (IWT),
Belgian Federal Science Policy Office (Belspo);
University of Oxford, United Kingdom;
Marsden Fund, New Zealand;
Japan Society for Promotion of Science (JSPS);
the Swiss National Science Foundation (SNSF), Switzerland;
A.~Gro{\ss} acknowledges support by the EU Marie Curie OIF Program;
J.~P.~Rodrigues acknowledges support by the Capes Foundation, Ministry of Education of Brazil.

We would like to thank G. Fogli, H. T. Janka, P, Mertsch, B. M\"uller, G.G. Raffelt, K. Sumiyoshi, I. Tamborra, and R. Tom\`{a}s for 
providing supernova model data and for helpful discussions. 
\end{acknowledgements}


\begin{thebibliography}{}

\bibitem[Abbasi et al.(2009)]{daq}
Abbasi, R., et al. 2009, Nucl. Instrum. Meth. A601, 294

\bibitem[Abbasi et al.(2010)]{ic3:pmt-paper}
Abbasi, R., et al. 2010, Nucl. Instrum. Meth. A618,  139

\bibitem[Abbasi et al.(2008)]{solar}
Abbasi, R., et al. 2008, Astrophys. J. Lett. 689, 65

\bibitem[Achterberg et al.(2006)]{firstyear}
Achterberg, A., et al. 2006, Astropart. Phys., 26, 155

\bibitem[Ackermann et al.(2006)]{iceproperties06}
Ackermann, M., et al. 1996, J. Geophys. Res., 111, D13203

\bibitem[Aglietta et al.(2002)]{IlNuovoCimentoA105.1793}
Aglietta, M., et al. 1992, Il Nuovo Cimento, A105, 1793

\bibitem[Ahlers et al.(2009)]{Ahlers}
Ahlers, M., Mertsch, P., Sakar, S. 2009, \prd, 80, 123017
 
\bibitem[Ahrens et al.(2002)]{bib:AMANDAold} 
Ahrens, J., et al. 2002, Astropart. Phys., 16, 345

\bibitem[Ahrens et al.(2004)]{pdd}
Ahrens, J., et al. 2004, IceCube Preliminary Design Document,\\
\url{http://www.icecube.wisc.edu/science/publications/pdd/}

\bibitem[Alekseev et al.(1987)]{JExpTheorPhys.45.589}
Alekseev, E. N., et al. 1987, J. Exp. Theor. Phys. Lett., 45, 589

\bibitem[Alimonti et al.(2009)]{NuclInstrMeth600.568}
Alimonti, G., et al. 2009, Nucl. Instrum. Meth., A600, 568

\bibitem[Amako et al.(2005)]{Amako}
Amako, K., et al. 2005, IEEE Trans. Nucl. Science, 52-4, 910

\bibitem[Antonioli et al.(2004)]{antonioli-2004-6}
Antonioli, P., et al. 2004, New J. Phys., 6, 114

\bibitem[Bahcall \& Piran(1983)]{Bahcall:1982fx}
Bahcall, J. N. \& Piran, T. 1983, \apj, 267, L77

\bibitem[Baum et al.(2011)]{bib:Baum}
Baum, V., et al. (IceCube Collaboration) 2011, in 32st Int. Cosmic Ray Conf., (Beijing, China: ICRC)

\bibitem[Beacom et al.(2001)]{BeacomBoydMezzacappa}
Beacom, J. F., Boyd, R. N., Mezzacappa, A.\& Vogel, P. 2001, \prd, 60, 73011

\bibitem[Bramall et al.(2005)]{Bramall}
Bramall, N.E. et al. 2005, Geophys. Res. Lett., 32, L21815

\bibitem[Brandt et al.(2011)]{Brandt}
Brandt, T Burrows, A., Ott, C. \& Livne, E. 2011, \apj, 728, 8

\bibitem[Bionta et al.(1987)]{PhysRevLett.58.1494}
Bionta, R. M., et al. 1987, \prl, 58, 1494

\bibitem[Buras et al.(2003)]{PhysRevLett.90.241101}
Buras, R., Rampp, M., Janka, H.-Th. \& Kifonidis, K. 2003, \prl, 90, 241101

\bibitem[Burrows et al.(1992)]{PhysRevD.45.3361}
Burrows, A., Klein, K. \& Gandhi, R. 1992, \prd, 45, 3361

\bibitem[Burrows \& Thompson(2002)]{Burrows:2002bz}
Burrows, A. \& Thompson, T. A. 2002, arXiv:astro-ph/0210212v1

\bibitem[Choubey et al.(2006)]{Choubey}
Choubey, S., Harries, N. P. \& ~Ross, G.G. 2006,
\prd, 74, 053010 

\bibitem[Dasgupta et al.(2010)]{Dasgupta-2010}
Dasgupta, B., et al. 2010, \prd, 81, 103005

\bibitem[Demiroers et al.(2011)]{bib:Demiroers}
Demiroers, L., Ribordy, M. \& ~Salathe, M. 2011, arXiv:astro-ph/1106.1937v2

\bibitem[Diehl et al.(2006)]{Diehl}
Diehl, R. et al. 2006, Nature, 439, 45

\bibitem[Dighe et al.(2004)]{dighe}
Dighe, A. S., et al. 2004, JCAP, 0401, 004

\bibitem[Eastman et al.(1996)]{eastman}
Eastman, R. G., Schmidt, B. P. \& Kirshner, A. 1996, \apj, 466, 911

\bibitem[Feser(2004)]{FeserThesis}
Feser, T. 2004, Echtzeit-Suche nach Neutrinoausbr\"uchen von Supernovae mit dem AMANDA-II Detektor, PhD thesis, Johannes Gutenberg-Universit\"at Mainz (in German) 

\bibitem[Fischer et al.(2010)]{Fischer}
Fischer, T., Whitehouse, S. C., Mezzacappa, A., Thielemann, F.-K., \& Liebend\"orfer, M. 2009, A\&A, 517, A80

\bibitem[Fukuda et al.(2002)]{NuclInstrumMethA.501.418}
Fukuda, K. et al. 2002, Nucl. Instrum. Meth. A, 501, 418

\bibitem[Giunti \& Kim(2007)]{snrate}
Giunti, C. \& Kim, C. W. 2007, Fundamentals of neutrino physics and astrophysics (New York: Oxford University Press), 517

\bibitem[Halzen et al.(1996)]{PhysRevD.53.7359}
Halzen, F., Jacobsen, J. E. \& Zas, E.	1996, \prd, 53, 7359

\bibitem[Halzen \& Raffelt(2009)]{halzen+raffelt}
Halzen, F. \& Raffelt, G. G. 2009, Phys.Rev.D, 80, 087301

\bibitem[Hamamatsu(2007)]{hamamatsu}
Hamamatsu Photonics 2007,\\ \url{http://sales.hamamatsu.com/assets/pdf/catsandguides/PMT_handbook_v3aE.pdf}

\bibitem[Haxton(1987)]{Haxton}
Haxton, W. C. 1987, \prd, 36, 2283

\bibitem[Haxton(1999)]{HaxtonR}
Haxton, W. C. \& Robertson, R. G. H. 1999, Phys. Rev. C, 59, 515

\bibitem[Hirata et al.(1987)]{PhysRevLett.58.1490}
Hirata, K., et al. 1987, \prl, 58, 1490

\bibitem[Hirata et al.(1988)]{PhysRevD.38.448}
Hirata, K., et al. 1988, \prd, 38, 448

\bibitem[H\"udepohl et al.(2010)]{Huedepohl}
H\"udepohl, L., M\"uller, B., Janka, H.-T., Marek, A., \& Raffelt, G.G. 2010, 
\prl, 104, 251101

\bibitem[Ikeda et al.(2007)]{Ikeda}
Ikeda, M., et al. 2007, \apj, 669, 519 

\bibitem[Jacobsen(1996)]{jacobsen}
Jacobsen, J. E. 1996, Simulating the detection of muons and neutrinos in deep Antarctic ice (PhD thesis, University of Wisconsin-Madison)

\bibitem[Janka et al.(2006)]{janka}
Janka, H.-Th., et al. 2007, Phys. Rep., 442, 38

\bibitem[Kachelriess et al.(2005)]{PhysRevD.71.063003}
Kachelriess, M., Tom\`{a}s, R., Buras, R., Janka, H.-Th., Marek, A. \& Rampp, M. 2005, \prd, 71, 063003

\bibitem[Keil et al.(2003)]{keil}
Keil, M. T., Raffelt, G. G. \& Janka, H.-Th. 2003, \apj, 590, 971

\bibitem[Kitaura et al.(2006)]{AstrAstrop.450.1}
Kitaura, F. S., Janka, H.-Th. \& Hillebrandt, W. 2006, A\&A, 450 1, 345

\bibitem[Kolbe et al.(2002)]{Kolbe}
Kolbe, E., Langanke, K. \& Vogel, P. 2002, \prd, 66, 13007

\bibitem[Kotake et al.(2006)]{kotake}
Kotake, K., Sato K. \& Takahashi, K., Rep. Prog. Phys., 69, 971

\bibitem[Langanke et al.(1996)]{Langanke}
Langanke, K. H.,  Vogel, P. \& Kolbe, E. 1996,  \prl, 76, 2629 

\bibitem[Lattimer \& Swesty(1991)]{Lattimer}
Lattimer, J. \& Swesty, D. 1991,  Nucl. Phys. A, 535, 331

\bibitem[Lundberg et al.(2007)]{Lundberg:2007yg}
Lundberg, J., et al. 2007, Nucl. Instrum. Meth. A, 581, 619

\bibitem[Marciano \& Parsa(2003)]{marciano:03}
Marciano, W. J. \& Parsa, Z. 2003, J. Phys. G: Nucl. Part. Phys., 29, 2629

\bibitem[Marek et al.(2009)]{Marek}
Marek, A., Janka,  H.-T. \& M\"uller, E. 2009, A\&A, 496 2, 475 

\bibitem[Meyer(2010)]{HOMeyer2010}
Meyer, H. O. 2010, Europhys. Lett., 89, 58001 

\bibitem[Mirizzi et al.(2006)]{Mirizzi}
Mirizzi, A., Raffelt, G. G.\& Serpico, P. D. 2006,  JCAP, 0605, 012

\bibitem[Novoseltseva et al.(2009)]{Novoseltseva}
Novoseltseva, R. V. et al. (Baksan Collaboration) 2009, in 31st Int. Cosmic Ray Conf., (Lodz, Poland: ICRC)

\bibitem[Pratt et al.(1999)]{IRIDIUM}
Pratt, S. R., Raines, R. A., Fossa Jr., C. E. \& Temple, M. A. 1999, IEEE Commun. Surveys Tuts., 2, 2

\bibitem[Pagliaroli et al.(2009a)]{Pagliaroli-a}
Pagliaroli, G., Vissani, F., Constantini, M. L. \& Ianni, A. 2009, \prl, 104, 031102

\bibitem[Pagliaroli et al.(2009b)]{Pagliaroli-b}
Pagliaroli, G., Vissani, F., Constantini, M. L. \& Ianni, A. 2009, Astropart. Phys., 31, 163

\bibitem[Price et al.(2002)]{PRICE}
Price, B., et al. 2002, Proc. Natl. Acad. Sci. USA, 99, 7844

\bibitem[Pryor et al.(1988)]{Pryor}
Pryor, C., Roos, C. \& Webster, M. 1988,  Astrophys. J., 329, 335

\bibitem[Richard(2008)]{Richardthesis}
Richard, A. S. 2008, A Study of the Sensitivity of the IceCube Detector to a Supernova Explosion 
(MSc thesis, Southern University, Baton Rouge, LA, USA)

\bibitem[Scheffler \& Elsasser(1998)]{Scheffler}
Scheffler, H. \& Elsasser, H. 1998, Physics of the Galaxy and Interstellar Matter (Berlin: Springer-Verlag)

\bibitem[Shen(1998)]{Shen}
Shen, H., Toki, H., Oyamatsu, K. \& Sumiyoshi, K. 1998, Nucl. Phys. A, 637, 435

\bibitem[Strom(1994)]{Strom}
Strom, R. G. 1994, A\&A, 288, L1

\bibitem[Strumia \& Vissani(2003)]{strumia}
Strumia, A. \& Vissani, F. 2003, Phys. Lett. B, 564, 42

\bibitem[Sumiyoshi et al.(2007)]{Sumiyoshi}
Sumiyoshi, K., Yamada S., \& Suzuki, H. 2007,  \apj, 667, 382  

\bibitem[Suzuki(1991)]{Suzuki}
Suzuki, H. (1991), Num. Astrophys. Japan, 2, 267

\bibitem[Takahara \& Sato(1988)]{Takahara}
Takahara M. \& Sato, K. 1988, Prog. Theor. Phys., 80, 861

\bibitem[Tom\`{a}s et al.(2004)]{Tomas}
Tom\`{a}s, R., et al. 2004, JCAP, 0409, 015

\bibitem[Thompson et al.(2003)]{thompson-2003-592}
Thompson, T. A., Burrows, A \& Pinto, P. A. 2003, \apj, 592, 434

\bibitem[Totani et al.(1997)]{AstropPhys.496.216}
Totani, T., Sato, K., Dalhed, H. E., Wilson, J. R. 1998, Astrop. Phys., 496, 216

\bibitem[Tilav et al.(2009)]{tilav_icrc}
Tilav, S. et al. (IceCube Collaboration) 2009, in 31st Int. Cosmic Ray Conf., (Lodz, Poland: ICRC)

\bibitem[Vissani \&  Paglioroli(2009)]{bib:Vissani}
Vissani, F. \& Paglioroli, G. 2009, Astronomy Letters, 35-1, 3

\bibitem[Vogel \& Beacom(1999)]{PhysRevD.60.053003}
Vogel, P. \& Beacom, J. F.	1999, \prd, 60, 053003

\bibitem[Zeitnitz et al.(1994)]{geantgcalor}
Zeitnitz, C. \& Gabriel, T. A. 1994, Nucl. Instrum. Meth. A, 349, 106

\end{thebibliography}
\end{document}